\pdfoutput=1

\documentclass[11pt]{article}

\usepackage[preprint]{acl}
\usepackage{times}
\usepackage{latexsym}
\usepackage{kotex}
\usepackage[T1]{fontenc}

\usepackage[utf8]{inputenc}

\usepackage{microtype}

\usepackage{inconsolata}

\usepackage{graphicx}

\usepackage{xcolor,colortbl}
\usepackage{lipsum}  
\usepackage{booktabs}
\usepackage{enumerate}
\usepackage{lscape}
\usepackage{amsfonts}
\usepackage{boxedminipage}

\newcommand{\goinguppp}{} 
\newcommand{\goingdown}{} 
\newcommand{\goingstay}{\checkmark} 
\newcommand{\goingrand}{} 
\newcommand{\rephrased}{} 
\newcommand{\checkedall}{}

\title{Examining Identity Drift in Conversations of LLM Agents}

\author{Junhyuk Choi, Yeseon Hong, Minju Kim  \and Bugeun Kim  \\
Department of Artificial Intelligence, 
Chung-Ang University\\
Seoul, Republic of Korea \\
\texttt{\{chlwnsgur129, ghddptjs, minjunim, bgnkim\}@cau.ac.kr} \\
}

\begin{document}
\maketitle

\begin{abstract}
Large Language Models (LLMs) show impressive conversational abilities but sometimes show identity drift problems, where their interaction patterns or styles change over time. As the problem has not been thoroughly examined yet, this study examines identity consistency across nine LLMs. Specifically, we (1) investigate whether LLMs could maintain consistent patterns (or identity) and (2) analyze the effect of the model family, parameter sizes, and provided persona types. Our experiments involve multi-turn conversations on personal themes, analyzed in qualitative and quantitative ways. Experimental results indicate three findings. (1) Larger models experience greater identity drift. (2) Model differences exist, but their effect is not stronger than parameter sizes. (3) Assigning a persona may not help to maintain identity. We hope these three findings can help to improve persona stability in AI-driven dialogue systems, particularly in long-term conversations.
\end{abstract}

\section{Introduction}

Recent research has actively explored the utilization of Large Language Models (LLMs) as chatbot systems by assigning them specific personas \citep{samuel2024personagym,nandkumar2024enhancing,tseng-etal-2024-two}. To enhance user satisfaction in such systems, maintaining the consistency of the persona assigned to the LLM is critical. If the persona of an LLM loses its consistency, it may fail to deliver the user experience expected by the users, leading to usability issues \citep{chatbot1}. So, researchers recently focused on investigating whether LLMs can preserve persona during a conversation, focusing on two aspects of persona: (1) memory that avoids conflict in conversation and (2) identity\footnote{Here, we refer to the term 'identity' as factors that influence LLMs responses, such as behavioral patterns or talking style. This differs from psychological identity or consciousness, which we believe LLMs do not have.} that maintains talking style or response patterns. Among the two aspects, we focus on whether LLMs can retain the given identity.



Regarding the identity of persona, existing studies focused on LLMs' identity \citep{psychobench,sotopia,llmcollabopsychology, llminteraction1} without any conversation. Mainly, most researchers examined which identity LLMs exhibit in a specific isolated situation. Though existing work revealed LLMs have a stable identity without any interaction, it is questionable whether LLMs can retain such identity throughout a long conversation. As many reports suggest that LLMs are very sensitive to contextual changes\citep{sclar2024quantifying}, so having a conversation may make an `identity drift' of LLMs during the interaction. A single case study on GPT \citep{llminteraction1} supports this claim: identity can be changed only with a few agent interactions. Despite the case study, the result cannot be easily generalized to other models due to the difference in model families and parameter sizes. Therefore, we need a study to identify model-specific effects on identity drift.

Thus, this paper compares the patterns of identity drift across nine LLMs and attempts to reveal the cause of such drifts. Especially, as our motivation begins with the persona of chatbots, we wanted to know whether LLMs suffer identity drifts during a conversation. In the experiment, we asked two LLM agents to discuss 36 themes that are related to one's life, emotions, values, and feelings. We borrowed these themes from human study \citep{aron1997experimental} since they make agents discuss their virtual identity. After collecting conversational logs, we analyze identity drift patterns with the following two questions.



\textit{RQ1. How do structural differences among LLMs affect identity drift?}

This research question focuses on the effect of model structure. As parameter sizes and model families may affect the performance and behavior of LLMs, we also suspect that such differences can cause changes in identity drifts. Thus, we employ a systematic comparison of identity patterns. Using topic modeling and PsychoBench \citep{psychobench}, we successfully identified a relationship between model structure and identity drift. Here, we decided not to provide a persona as input because the persona may introduce unwanted effects.

\textit{RQ2. How does the provided persona affect identity drift?}

We pose another research question to observe the effect of persona. Specifically, we provide two kinds of personas to LLMs regarding how much the prompt asks LLMs to be influenced by the conversational partner: low and high. As instruction-tuned LLMs try to follow inputs as instruction, we suspect that low-influence persona may show a lower identity drift than the others. So, we used LLMs, which showed strong drifts in RQ1, to test whether the effect of persona is larger than that of the model.

\section{Related Work}

Researchers have been examining two factors that affect consistency in conversations: memory and identity. Because people generally expect consistency throughout a dialogue, researchers first started by examining memory consistency, which can easily form a task. A large body of existing research has focused on how memory is retained, largely verifying whether an LLM continues to remember certain information during conversation \citep{tseng-etal-2024-two,chen2023learning,maharana-etal-2024-evaluating,zhang2024survey,afzoon2024persobench}. For instance, \citet{chen2023learning} analyzed how consistently an LLM can uphold a given memory. Meanwhile, \citet{maharana-etal-2024-evaluating} created the LoCoMo dataset to investigate how well they remember information over prolonged conversations. 

However, memory is not the only factor that affects task performance or the naturalness of a dialogue; identity should be provided \citep{autogen, metaagents,teacherStudentInteraction,llmcollabopsychology}. 
For example, \citet{llmcollabopsychology} assessed LLMs' ability to engage in cooperative interactions based on Society of Mind theory \citep{minsky1988society} in a multi-agent environment.
Similarly, \citet{teacherStudentInteraction} reported that it is possible to model a conversational question-answering task as a virtual interaction between a teacher agent and a student agent using an LLM. By qualitatively assessing the quality of the interaction, they found that providing two identities could improve the interaction process in a more human-like manner. 

Also, \citet{metaagents} simulated a job fair scenario with two agents: a job seeker and an employer. They explored how their cooperative interaction affects task performance. 
However, all of these studies assume that the identity remains unchanged when a conversation progresses. Considering that the memory of a persona changes during a conversation, the identity could also be changed.

Hence, recently, researchers attempted to quantify the identity of persona before measuring its consistency. Some researchers designed benchmarks measuring the identity of LLM \citep{psychobench,sotopia,llmcollabopsychology, llminteraction1}. 
For example, \citet{psychobench} assessed the identity of LLMs using fourteen types of questionnaires. Though they found that different LLMs exhibit different identities, they did not let LLMs converse before measuring the identity.
However, impact of conversation is crucial because accumulated chat histories can introduce unexpected effects, as memory-related studies suggested.
\citet{llminteraction1} supports this claim. They demonstrated that GPT models in an interaction setting tend to adopt one another's persona, failing to maintain identity. Though this paper addressed the problem we call identity drift, it has some limitations when applied to conversational agents; the interaction was unidirectional compared to a usual conversation, as they asked agents to continue to write others' work. We suspect that, in a bidirectional conversation, the tendency of identity drift may not be the same as in a unidirectional one. Therefore, it is yet unanswered whether LLMs can consistently maintain the identity of the given persona in a bidirectional conversation.

\section{Experiments}

To investigate factors influencing identity drift issue of LLMs, we conduct an experiment \footnote{Code is available at [blinded for review].}. The experiment asks two LLM agents discuss about 36 themes. During the conversation, we collect their conversation logs and measure identity based on the conversation. Using both qualitative and quantitative analyses, we attempt to answer two research questions about which factor may affect identity drift. Thus, in this section, we first describe LLM agents used (Sections \ref{sec:exp-models} and \ref{sec:exp-identity}). Next, we describe how we let agents generate a conversation (Section \ref{sec:exp-procedure}). We also illustrate our qualitative and quantitative analysis methods (Sections \ref{sec:exp-topic} and \ref{sec:exp-stat}).

\subsection{RQ1: Language Models Tested}
\label{sec:exp-models}

For RQ1, we compared nine models, considering their popularity, parameter size, and architecture. Based on popularity, we selected GPT, the most famous black box LLM, and three famous open-sourced families: LLaMA, Mixtral, and Qwen. Table \ref{tab:models} shows the nine models with their parameter sizes\footnote{We assigned Mixtral models by their active parameter sizes (13B and 39B), according to \url{https://mistral.ai/en/news/mixtral-8x22b}.}. According to parameter sizes, we partitioned open-sourced models into three categories: small (models with $<$ 20 billion parameters), medium (models with $<$ 100 billion parameters), and large (models with $\ge$ 100 billion parameters). This categorization allows a systematic comparison of performance and model characteristics based on parameter scale. We did not assigned GPT models into any size groups since OpenAI did not officially disclose the parameter size of the GPT family. To focus on the effect of model itself, it is worth noting that we did not provide any identity-related information in the input prompt.

\begin{description}
    \item[GPT] This family comprises GPT-3.5 Turbo \citep{gpt3} and GPT-4o \citep{hurst2024gpt}. Although their parameter sizes remain undisclosed, these models were included in the experiment due to their high performance and widespread recognition in practice.
    \item[LLaMA3.1] This family includes LLaMA 3.1-8B, 3.1-70B, and 3.1-405B \citep{llama3}. While sharing the same basic architecture, they differ substantially in parameter size. Note that LLaMA provides one model with the largest parameter size.
    \item[Mixtral] This family contains Mixtral8x7B and Mixtral8x22B \citep{mixtral}. It employs a Mixture-of-Experts (MoE) architecture, which differs from other two open-sourced families. Thus, comparing Mixtral and others can prompt probing of how MoE influences potential identity shifts and the resulting conversation.
    \item[Qwen] This family encompasses Qwen2 7B and Qwen2 72B \citep{qwen2}. Advertised as particularly adept at conversational tasks, these models were considered suitable for analyzing how model identity drifts through extended interactions.
\end{description}

\begin{table}
    \centering
    \begin{tabular}{l|rrr}
    \toprule
        Family & \multicolumn{3}{c}{Parameter Sizes}\\
        \cmidrule(lr){2-4}
        & Small & Medium & Large \\
        \midrule
        LLaMA 3.1 &   8B & 70B & 405B \\
        Mixtral & 8x7B & 8x22B & \\
        Qwen 2 & 7B & 72B & \\
        \midrule
        GPT & \multicolumn{3}{c}{\textit{Undisclosed}: 3.5 Turbo, 4o}\\
        \bottomrule
    \end{tabular}
    \caption{Models tested in our experiment}
    \label{tab:models}
\end{table}


\subsection{RQ2: Providing identity}
\label{sec:exp-identity}


After investigating RQ1, we examine the effect of the provided persona. As we suspect the effect of persona is not large enough to offset the effect of model-related factors, we used two LLMs whose identity drifts are the most severe among the nine models. Though users expect LLMs can maintain consistent identity, those two models should maintain the identity to meet the expectation.

Also, we set two types of identity, regarding how the description instructs the model. As those nine LLMs are trained to follow instructions, the result may be affected by how the persona is influenced by the others. Thus, we suspect that LLMs may suffer more identity drifts when we provide an identity highly influenced. So, we define two groups: (1) \textit{high-influence} group and (2) \textit{low-influence} group. High-influence personas have emotionally sensitive and empathetic identity, thereby allowing for more flexible changes in their response and identity during the conversation. In contrast, we set low-influence personas as outgoing and goal-oriented, which are not directly related to emotional sensitivity. Detailed information on these personas can be found in the Appendix. We created 20 identities for each group. Note that we also provided the basic information of the persona (e.g., name, gender, and age) to mirror the usual usecase of persona-provided chatbots.

\subsection{Procedure for Generating conversation}
\label{sec:exp-procedure}



Our generation procedure is inspired by a psychological study \citep{aron1997experimental}. We chose the study because of two reasons. First, the method suggests a scientific way to identify changes during a conversation. They let humans have a conversation about 36 themes and measured human psychological states three times within the conversation. By comparing three measured values, they could statistically identify the changes in human states. As we also aimed to measure changes in identity, we borrowed their experimental setup. Second, the method uses materials that are highly related to identity of someone. The 36 themes used in the study directly or indirectly ask participants to answer their thoughts about their lives, values, or motivations. So, it is highly likely that the answer contains the related concepts about their identity. In the view of LLMs, such answers may ignite some related tokens during the generation procedure. That is, the identity may be easily affected by the words in the previous discussion. Thus, we adopted the study. 

In the generation procedure, we asked two agents answer the 36 themes in \citet{aron1997experimental}. For each theme, we pose a question about the theme. One of the agents generates a response to the question, considering previous conversational history. Then, the other agent generates response to the question, considering the first agent's answer and previous history. We repeated this procedure until the end of 36 themes and collected conversation logs to answer research questions. For RQ1, we simulated 20 conversations for each LLM. For RQ2, we simulated 10 conversations for each persona group: we paired similar personas to avoid the identity drift effect reported by \citep{llminteraction1}. To obtain diverse conversation logs and mirror the real-world usage, we set the temperature parameter at 0.7\footnote{This value was the default temperature value when we experimented. Though the default value changed to 1.0, we believe that such a difference may not severely harm our experimental result.}. Consequently, we gathered 400 logs for each research question.

\subsection{Qualitative: Topic modeling}
\label{sec:exp-topic}

As a qualitative analysis, we employed BERTopic \citep{grootendorst2022bertopic} which is a topic modeling method. The unit of analysis for the topic exploration was a single utterance, defined as one participant’s response to one of the 36 themes. Notably, we included only generated answers, excluding any statements or prompts provided to the LLM participants. Given that there were 20 conversations with two participants per session, each LLM generated 1,440\footnote{1440 = 20 × 2 × 36} utterances. To obtain more meaningful topics, we removed stop-words, used an English-based embedding, and set the minimum topic size as 50. 

To answer two research questions, we identified topics for each condition and compared across conditions. We believe comparing differences in topic analysis results may provide insights about differences in conditions. For example, we ran topic modeling for three times for parameter size groups: small, middle and large. Similarly, we ran topic modeling for four times for model families: GPT, LLaMA, Mixtral, and Qwen. Also, we separately extracted topics for high-influenced and low-influenced identities for RQ2. We chose the ten most representative topics from each run, and associated topics with one of the 36 themes. After that, we compared representative words among conditions to find the differences between them.

\begin{table}[!t]
    \small
    \setlength{\tabcolsep}{2pt}
    \centering
    \begin{tabular}{cp{.8\columnwidth}|c}
    \bottomrule
    \multicolumn{2}{l}{\cellcolor{gray!20} Small-sized open-source models ($\le$ 10B)} & {\cellcolor{gray!20}Theme}\\
    \toprule
    \#0 & \textbf{friendship}, trust, respect, mutual, \textbf{means} & 20\\
    \#1 & \textit{users}, language, accomplishments, accomplishment, \textit{assist} & (AI) \\
    \#2 & \textbf{feel}, way, appreciate, grateful, admire & 31\\
    \#3 & \textbf{regret}, \textbf{told}, expressing, \textbf{having}, feelings & 33\\
    \#4 & dont, \textit{digital}, exist, existence, designed & (AI) \\
    \#5 & shared, understanding, conversations, mutual, deep & 20\\
    \#6 & death, living, live, \textbf{die}, \textbf{hunch} & 7\\
    \#7 & rehearsing, \textbf{rehearse}, ensure, helps, especially & 3\\
    \#8 & humor, topics, \textbf{jokes}, issues, sensitive & 32\\
    \#9 & singing, sang, \textbf{sing}, karaoke, fun & 5\\
    
    \bottomrule
    \multicolumn{2}{l}{\cellcolor{gray!20} Middle-sized open-source models (10B - 100B)} & {\cellcolor{gray!20} Theme}\\
    \toprule
     \#0 & way, really, appreciate, \textbf{feel}, qualities & 31\\
     \#1 & know, \textbf{friendship}, honesty, value, want & 20\\
     \#2 & statements, \textbf{shared}, \textbf{value}, growth, \textbf{conversations} & 25\\
     \#3 & \textbf{regret}, \textbf{told}, \textbf{having}, loved, \underline{ive} & 33\\
     \#4 & languages, \textbf{ability}, cultures, language, speak & 12\\
     \#5 & living, \textbf{die}, focusing, present, healthy & 7\\
     \#6 & \textbf{childhood}, \textbf{family}, happy, \textbf{warm}, \textbf{close} & 23 \\
     \#7 & fascinating, conversation, choose, elon, musk & 1\\
     \#8 & \textbf{accomplishment}, \textbf{greatest}, hard, proud, achievement & 15\\
     \#9 & \textbf{mother}, \textbf{relationship}, \underline{shes}, guidance, loving & 24\\
    \bottomrule
    
    \multicolumn{2}{l}{\cellcolor{gray!20} Large-sized open-source models ($>$ 100B)} & {\cellcolor{gray!20} Theme}\\
    \toprule
     \#0 & statements, \textbf{friendship}, life, having, grateful & 20\\
     \#1 & \underline{ive}, \textbf{accomplishment}, \textbf{life}, \textbf{greatest}, encouraged & 11\\
     \#2 & really, way, \underline{youre}, \textbf{feel}, \textbf{like} & 31\\
     \#3 & \textbf{regret}, \textbf{told}, \textbf{having}, \underline{ive}, think & 33\\
     \#4 & live, left, focus, try, make & 19\\
     \#5 & \textbf{feeling}, \underline{ive}, \underline{youre}, problem, advice & 36 \\
     \#6 & \textbf{embarrassing}, memory, ended, \textbf{moment}, painful & 29 \\
     \#7 & \textbf{affection}, \textbf{love}, relationship, mother, believe & 21\\
     \#8 & \underline{id}, able, famous, \textbf{ability}, language & 12\\
     \#9 & \textbf{know}, want, \underline{im}, \underline{id}, bit & 27\\
    \bottomrule
    \end{tabular}
    \caption{Top 10 topics discovered per parameter size groups. Underlined words are related to pronouns.}
    \label{tab:topic-size}
\end{table}

\renewcommand{\arraystretch}{0.5}
\begin{table*}
    \centering
    \begin{tabular}{l@{}r@{\;}|@{\;}c@{\;}c@{\;}|@{\;}c@{\;}c@{\;}c@{\;}|@{\;}c@{\;}c@{\;}|@{\;}c@{\;}c@{\;}||@{\;}c@{\;}c@{\;}|@{\;}c@{\;}c}
    
    \toprule
     & {\small Conditions:} & \multicolumn{9}{c||@{\;}}{\textit{Without} providing persona} & \multicolumn{4}{c}{\textit{With} persona} \\
    \cmidrule(r){3-11}\cmidrule(r){12-15}
     & {\small Family:} & \multicolumn{2}{c|@{\;}}{GPT} &  \multicolumn{3}{c|@{\;}}{LLaMA 3.1} & \multicolumn{2}{c|@{\;}}{Mixtral} & \multicolumn{2}{c||@{\;}}{Qwen 2} & \multicolumn{2}{c|@{\;}}{GPT-4o} & \multicolumn{2}{c}{L 405B} \\
    \cmidrule(r){3-4}\cmidrule(r){5-7}\cmidrule(r){8-9}\cmidrule(r){10-11}\cmidrule(r){12-13}\cmidrule(r){14-15}
    & & \texttt{\small 3.5T} & \texttt{\small \ 4o\ } & \texttt{\small \ 8B\ } & \texttt{\small 70B\ } & \texttt{\small 405B} & \texttt{\small \ 7B\ } & \texttt{\small 22B\ } & \texttt{\small \ 7B\ } & \texttt{\small 72B\ } 
    & \texttt{\small low} & \texttt{\small \ high} & \texttt{\small low} & \texttt{\small \ high} \\
    \midrule
    
    \multicolumn{15}{@{}l}{\large \textbf{(1) Personality}} \\
    \midrule
    {\small BFI} & {\small \textit{Openness}} & \goinguppp & \goingdown & \goinguppp & \goingstay & \goingdown & \goingstay & \goingstay & \goingstay & \goinguppp & \goingdown & \goingdown & \goingstay & \goingdown \\
     & {\small \textit{Conscientiousness}} & \goinguppp & \goingdown & \goinguppp & \goingstay & \goingdown & \goingstay & \goingstay & \goingstay & \goingstay &\goingdown & \goingdown & \goingstay & \goingdown \\
     & {\small \textit{Extraversion}} & \goingrand & \goingdown & \goinguppp & \goingdown & \goingstay & \goingstay & \goingdown & \goingstay & \goingstay &\goingdown & \goingdown & \goingstay & \goingdown \\
     & {\small \textit{Agreeableness}} & \goinguppp & \goingdown & \goingstay & \goingstay & \goingdown & \goingstay & \goingstay & \goingstay & \goingstay &\goingdown & \goingdown & \goingdown & \goingdown \\
     & {\small \textit{Neuroticism}} & \goinguppp & \goingdown & \goinguppp & \goingstay & \goingdown & \goingstay & \goingdown & \goingstay & \goingstay &\goingdown & \goingdown & \goingstay & \goingstay\\
    \midrule
    {\small EPQ-R} & {\small \textit{Extraversion}} & \goinguppp & \goingdown & \goingdown & \goingdown & \goingdown & \goingdown & \goingstay & \goingstay & \goingstay &\goingdown & \goingdown & \goingdown & \goingdown \\
     & {\small \textit{Psychoticism}} & \goinguppp & \goingdown & \goingdown & \goingdown & \goingdown & \goingdown & \goingstay & \goingstay & \goingstay &\goingdown & \goingdown & \goingdown & \goingdown \\
     & {\small \textit{Neuroticism}} & \goinguppp & \goingdown & \goingdown & \goingdown & \goingdown & \goingdown & \goingstay & \goingstay & \goingstay &\goingdown & \goingdown & \goingdown & \goingdown \\
     & {\small \textit{Lying}} & \goinguppp & \goingdown & \goingstay & \goingdown & \goingdown & \goingdown & \goingstay & \goingstay & \goingstay &\goingdown & \goingdown & \goingdown & \goingdown  \\
    \midrule
    {\small DTDD} & {\small \textit{Machiavellianism}} & \goingdown & \goingdown & \goingdown & \goingdown & \goingdown & \goingdown & \goingdown & \goinguppp & \goingstay &\goingdown & \goingstay & \goingstay & \goingdown\\
     & {\small \textit{Psychopathy}} & \goingdown & \goingdown & \goingstay & \goingdown & \goingdown & \goingstay & \goingdown & \goingstay & \goingstay &\goingdown & \goingstay & \goingstay & \goingdown  \\
     & {\small \textit{Narcissism}} & \goingdown & \goingrand & \goingstay & \goingdown & \goingdown & \goingstay & \goingdown & \goingstay & \goingstay &\goingdown & \goingstay & \goinguppp & \goingdown \\
    \midrule
    & {\small \textit{Total count (12)}} & 0 & 0 & 4 & 4 & 1 & 7 & 7 & 11 & 11 & 0 & 3 & 6 & 1 \\
    \multicolumn{15}{c}{\phantom{M}}\\
    \multicolumn{15}{@{}l}{\large \textbf{(2) Interpersonal Relationship}}\\
    \toprule
    {\small BSRI} & {\small \textit{Masculine}} & \goinguppp & \goingstay & \goingstay & \goingdown & \goingdown & \goingstay & \goingdown & \goingstay & \goingstay &\goingrand & \goingdown & \goingdown & \goingstay \\
     & {\small \textit{Feminine}} & \goinguppp & \goingdown & \goinguppp & \goingdown & \goingdown & \goingstay & \goingstay & \goingstay & \goinguppp &\goingrand & \goingdown & \goingdown & \goingstay \\
    \midrule
    {\small CABIN} & {\small \textit{Realistic}} &  \goingstay & \goingdown & \goingstay & \goingdown & \goingdown & \goingstay & \goingstay & \goingdown & \goingdown &\goingdown & \goingdown & \goingstay & \goingdown \\
     & {\small \textit{Investigate}} & \goingstay & \goingrand & \goingstay & \goingdown & \goingstay & \goingstay & \goingstay & \goingdown & \goingdown &\goingdown & \goingdown & \goingstay & \goingdown \\
     & {\small \textit{Artistic}} & \goinguppp & \goingstay & \goingstay & \goingdown & \goingdown & \goingstay & \goingstay & \goingdown & \goingdown &\goingdown & \goingdown & \goingstay & \goingdown \\
     & {\small \textit{Social}} & \goingstay & \goinguppp & \goingstay & \goingdown & \goingdown & \goingstay & \goingstay & \goingdown & \goingdown &\goingdown & \goingdown & \goingstay & \goingdown \\
     & {\small \textit{Enterprising}} & \goingstay & \goingstay & \goingstay & \goingdown & \goingdown & \goingstay & \goingstay & \goingdown & \goingdown &\goingdown & \goingdown & \goingstay & \goingdown \\
     & {\small \textit{Conventional}} & \goingstay & \goingrand & \goingstay & \goingdown & \goingdown & \goingstay & \goingstay & \goingdown & \goingdown &\goingdown & \goingdown & \goingstay & \goingdown \\
    \midrule
    {\small ICB} & {\small \textit{Overall}} & \goingdown & \goingstay & \goinguppp & \goinguppp & \goingstay & \goingstay & \goingstay & \goingstay & \goinguppp &\goingstay & \goingdown & \goingstay & \goingstay \\
    \midrule
    {\small ECR-R} & {\small \textit{Attachment Anxiety}} & \goingstay & \goingdown & \goingstay & \goingdown & \goingdown & \goingdown & \goingdown & \goinguppp & \goingstay &\goingdown & \goingstay & \goingdown & \goingrand \\
     & {\small \textit{Attachment Avoidance}} & \goinguppp & \goingdown & \goingstay & \goingdown & \goingdown & \goingstay & \goingstay & \goingrand & \goingstay &\goingdown & \goingstay & \goingdown & \goingdown\\
     \midrule
    {\small MFQ} & {\small \textit{Stimulating companionship}} & \goingdown & \goinguppp & \goingstay & \goinguppp & \goinguppp & \goingstay & \goinguppp & \goingstay & \goinguppp &\goinguppp & \goinguppp & \goinguppp & \goinguppp \\
     & {\small \textit{Help}} & \goingdown & \goinguppp & \goingstay & \goinguppp & \goinguppp & \goingstay &  \goinguppp & \goingstay & \goinguppp &\goinguppp & \goinguppp & \goinguppp & \goinguppp \\
     & {\small \textit{Intimacy}} & \goingdown & \goinguppp & \goingstay & \goinguppp & \goinguppp & \goingstay & \goinguppp & \goingstay & \goinguppp &\goinguppp & \goinguppp & \goinguppp & \goinguppp \\
     & {\small \textit{Reliable alliance}} & \goingdown & \goinguppp & \goingstay & \goinguppp & \goinguppp & \goingstay & \goinguppp  & \goingstay & \goinguppp &\goinguppp & \goinguppp & \goinguppp & \goinguppp \\
     & {\small \textit{Self-validation}} & \goingdown & \goinguppp & \goingstay & \goinguppp & \goinguppp & \goingstay & \goinguppp & \goingstay & \goinguppp &\goinguppp & \goinguppp & \goinguppp & \goinguppp \\
     & {\small \textit{Emotional security}} & \goingdown & \goinguppp & \goingstay & \goinguppp & \goinguppp & \goingstay & \goinguppp & \goinguppp & \goinguppp &\goinguppp & \goinguppp & \goinguppp & \goinguppp \\
    \midrule
    & {\small \textit{Total count (17)}} & 6 & 4 & 15 & 0 & 2 & 16 & 9 & 8 & 3 & 1 & 2 & 7 & 3 \\
    \multicolumn{15}{c}{\phantom{M}}\\
    \multicolumn{15}{@{}l}{\large \textbf{(3) Motivation}}\\
    \toprule
    {\small GSE} & {\small \textit{Overall}} & \goingstay & \goingstay & \goinguppp & \goingrand & \goingstay & \goingrand & \goingstay & \goingrand & \goingdown & \goinguppp & \goingdown & \goinguppp & \goingdown \\
    \midrule
    {\small LOT-R} & {\small \textit{Overall}} & \goingrand & \goingrand & \goinguppp & \goinguppp & \goingstay & \goingstay & \goingstay & \goinguppp & \goinguppp & \goingstay & \goingdown & \goinguppp & \goingdown \\
    \midrule
    {\small LMS} & {\small \textit{Rich}} & \goinguppp & \goingdown & \goingstay & \goingdown & \goingdown & \goingdown & \goingdown & \goingrand & \goingstay & \goingdown & \goingdown & \goingdown & \goingdown \\
     & {\small \textit{Motivator}} & \goingdown & \goingdown & \goinguppp & \goingdown & \goingdown & \goingdown & \goingdown & \goingdown & \goinguppp & \goingdown & \goingdown & \goingstay & \goingdown \\
     & {\small \textit{Important}} & \goingdown & \goingdown & \goinguppp & \goingdown & \goingdown & \goingdown & \goingdown & \goingrand & \goingstay & \goingdown & \goingdown & \goingstay & \goingdown \\
    \midrule
    & {\small \textit{Total count (5)}} & 1 & 1 & 1 & 0 & 2 & 1 & 2 & 0 & 2 & 1 & 0 & 2 & 0 \\
    \multicolumn{15}{c}{\phantom{M}}\\
    \multicolumn{15}{@{}l}{\large\textbf{(4) Emotion}}\\
    \toprule
    {\small EIS} & {\small \textit{Overall}} & \goingrand & \goingdown & \goingstay & \goingdown & \goingdown & \goingstay & \goingdown & \goingdown & \goingdown & \goingrand & \goingdown & \goingdown & \goingstay \\
    \midrule
    {\small WLEIS} & {\small \textit{Self-emotion appraisal}} & \goingrand & \goingdown & \goingstay & \goingstay & \goinguppp & \goingstay & \goinguppp & \goingdown & \goingrand & \goingrand & \goingdown & \goinguppp & \goingstay \\
     & {\small \textit{Others' emotion appraisal}} & \goingrand & \goingdown & \goinguppp & \goingstay & \goinguppp & \goinguppp & \goinguppp & \goingdown & \goingrand & \goingrand & \goingdown & \goinguppp & \goingstay \\
     & {\small \textit{Use of emotion}} & \goingrand & \goingdown & \goinguppp & \goingdown & \goinguppp & \goinguppp & \goinguppp & \goinguppp & \goingrand & \goingrand & \goingdown & \goinguppp & \goingstay \\
     & {\small \textit{Regulation of emotion}} & \goingrand & \goingdown & \goinguppp & \goingdown & \goingstay & \goingstay & \goinguppp & \goingrand & \goingrand & \goingrand & \goingdown & \goinguppp & \goingstay \\
    \midrule
    {\small Empathy} & {\small \textit{Overall}} & \goingrand & \goingdown & \goingstay & \goingdown & \goingstay & \goingdown & \goingstay & \goingstay & \goingrand & \goingdown & \goingstay & \goingstay & \goingstay \\
    
    \midrule
    & {\small \textit{Total count (6)}} & 0 & 0 & 3 & 2 & 2 & 3 & 1 & 1 & 0 & 0 & 1 & 1 & 6\\
    \end{tabular}
    
    \caption{Verification of whether the identity of persona was retained during the conversation for each subscale. Checkmarks (\checkmark) indicate the identity change is statistically insignificant in both Friedman and posthoc tests. Detailed statistical results are shown in Appendix (Tables from \ref{tab:app-full-1} to \ref{tab:qwen-comparison}).}
    \label{tab:result}
\end{table*}
\renewcommand{\arraystretch}{1.0}

\subsection{Quantitative: PsychoBench and MFQ}
\label{sec:exp-stat}

As a quantitative analysis, we adopted PsychoBench \citep{psychobench} and Mcgill's Friendship Questionnaire (MFQ; \citet{McGill}). These artifacts can measure identity of persona. PsychoBench contains thirteen questionnaires from psychology, quantifying four parts of one's identity: personality, interpersonal relationship, motivation, and emotion. We expect these four parts keep unchanged during a conversation. MFQ quantifies how one thinks about the conversational partner. We included this questionnaire to track how the conversational agents think each other. Detailed descriptions for those fourteen questionnaires are in Appendix \ref{app:quest}.

We measured those questionnaires three times within a conversation. Inspired by \citet{aron1997experimental}, we set three snapshots for each conversation log: after answering 12th, 24th, and 36th themes. Then, we applied PsychoBench and MFQ on those snapshots. As in PsychoBench, we asked LLMs to answer the questionnaire ten times with temperature zero to account for the primacy effect \citep{wang2023primacy}. Meanwhile, our method differs from PsychoBench in that we fed previous conversation logs to measure the identity based on the generated conversation logs. As a result, we can collect scored responses for each snapshot.

Using the scored responses, we performed statistical tests to identify identity drifts. First, we verify whether the identity changed on some snapshots. We used the repeated measure ANOVA or a Friedman tests \citep{anova, friedman}, regarding normality of scored responses. Second, we ensure consistency by checking pairwise post-hoc tests. We used Tukey's test or Wilcoxon signed-ranked test \citep{tukey, wilcoxon}, respectively. To mitigate potential Type I errors arising from multiple comparisons, we used Bonferroni correction to adjust p-values conservatively in the Wilcoxon test \citep{bonferroni}.

\section{Result and Discussion}

In this section, we summarize the experimental results in terms of the research questions. We first discuss qualitative and quantitative results of RQ1. Then, we illustrate the tendency we found in RQ2.

\subsection{RQ1: Effect of Structure}

The experimental result for RQ1 indicates that the effect of model-related factor exists. Specifically, parameter sizes showed a large impact on consistency. The effect of model family is relatively low, compared to the size.

\begin{table*}[!t]
    \small
    \setlength{\tabcolsep}{2pt}
    \centering
    \begin{tabular}{cp{.8\columnwidth}|c|cp{.8\columnwidth}|c}
    \bottomrule
    \multicolumn{2}{c}{\cellcolor{gray!20} GPT family} & {\cellcolor{gray!20} Theme} & 
    \multicolumn{2}{c}{\cellcolor{gray!20} LLaMA 3.1 family} & {\cellcolor{gray!20} Theme}\\
    \toprule
    \#0 & thoughtful, admire, genuine, appreciate, empathy & 28
    & \#0 & dont, personal, information, \textit{assist}, provide & (AI) \\
    \#1 & enjoy, value, meaningful, growth, appreciate & 8
    & \#1 & desire, value, nature, conversations, based &25\\
    \#2 & value, friendship, honesty, \textbf{important}, trust & 27
    & \#2 & way, really, feel, \underline{youre}, \textbf{like} & 31\\
    \#3 & \textbf{regret}, \textbf{told}, expressing, feelings, telling & 33
    & \#3 & \textbf{regret}, \textbf{told}, \textbf{having}, \underline{ive}, ones & 33\\
    \#4 & \underline{youd}, discuss, free, like, \underline{im} & (AI) 
    & \#4 & \textbf{famous}, \underline{id}, author, music, renowned & 2\\
    \#5 & \textbf{affection}, \textbf{love}, emotional, \textbf{play}, belonging & 21
    & \#5 & \textbf{friendship}, \underline{means}, having, accepts, connection & 20\\
    \#6 & \textbf{greatest}, \textbf{accomplishment}, far, completing, overcoming & 15
    & \#6 & \textbf{rehearse}, helps, avoid, ensure, yes & 3\\
    \#7 & \textbf{ability}, choose, \textbf{wake}, \textbf{tomorrow}, speak & 12
    & \#7 & da, leonardo, vinci, facinating, art & 1\\
    \#8 & \textbf{year}, \textbf{knew}, focus, left, prioritize & 19
    & \#8 & \underline{singing}, \underline{sang}, favorite, driving, ago & 5\\
    \#9 & \textbf{means}, \textbf{friendship}, having, trust, mutual & 20
    & \#9 & topics, \textbf{joked}, humor, issues, hurtful & 32\\

    \bottomrule
    \multicolumn{2}{c}{\cellcolor{gray!20} Mixtral family} & {\cellcolor{gray!20} Theme} 
    & \multicolumn{2}{c}{\cellcolor{gray!20} Qwen family} & {\cellcolor{gray!20} Theme}\\
    \toprule
     \#0 & appreciate, admire, humor, feel, kindeness & 31 
     & \#0 & \textit{ai}, dont, \textit{users}, \textit{assist}, information& (AI)\\
     \#1 & live, \textbf{living}, make, time, \textbf{die} &19 
     & \#1 & kindness, qualities, admire, humor, thoughtful& 31 \\
     \#2 & \textbf{told}, \textbf{regret}, expressing, \textbf{having}, express & 33 
     & \#2 & living, focusing, time, experiences, death& 7\\
     \#3 & \textbf{accomplishment}, \textbf{greatest}, \textbf{life}, career, work & 11, 15 
     & \#3 & impact, world, accomplishment, positive, career & 13 \\
     \#4 & \textbf{statements}, shared, value, importance, enjoy & 25 
     & \#4 & shared, interests, committed, \textbf{statements}, learning & 25 \\ 
    \#5 & \textit{users}, language, \textit{model}, \textit{artificial}, \textit{ai}& (AI) 
    & \#5 & \textbf{regret}, expressing, gratitude, feelings, loved& 33\\
    \#6 & humor, topics, mindful, \underline{jokes}, \underline{joking} & 32 
    & \#6 & honesty, respect, \textbf{friendship}, mutual, \textbf{value} & 16\\
    \#7 & \textbf{dinner}, obama, michelle, \textbf{guest}, \underline{choice} & 1 
    & \#7 & loss, \textbf{disturbing}, losing, profoundly, profound& 35 \\
    \#8 & \textbf{day}, \textbf{perfect}, relaxation, involve, activities & 4 
    & \#8 & languages, cultures, exposure, \textbf{ability}, different & 12\\
    \#9 & \textbf{mind}, \textbf{body}, mental, \underline{30yearold}, \textbf{retain} & 6 
    & \#9 & \underline{memories}, \textbf{treasured}, cherished, sharing, \textbf{memory} & 17\\
    \bottomrule
    \end{tabular}
    \caption{Top 10 topics discovered per family. Bold-faced words seem to be copied from the corresponding theme.}
    \label{tab:topic-family}
\end{table*}

\paragraph{Effect of parameter sizes}


According to the qualitative analysis, two notable changes were observed in the representative topics among different parameter sizes: those pertaining to ``AI" and to ``pronouns.'' The result is shown in Table \ref{tab:topic-size}. First, regarding AI, small LLMs refuse to engage in conversations on a given theme as they are an AI. As shown in Topics \#1 and \#4 for the small models, they tended to refuse or guard their own responses. This tendency was not observed in the medium or large models. So, though the safeguard was activated during the conversation in small models, that of middle or large models was not activated.

Second, regarding pronouns, large LLMs generates its responses based on fictitious information about itself or the other participant. Though pronouns are filtered by stop-words, there are some pronoun-based forms unfiltered by stop-word dictionary; for example, ``I've.'' Compared to the small models (0 pronouns), medium and large models (2 and 8 pronouns) have relatively high number of pronouns in the topic words. Due to the recency effect and other biases, such fictitious contents may influence subsequent conversations. This claim is supported by themes co-ocurring across size groups. For example, Theme 31 asks about one's perception of the other participant, and only the large models used second-person pronouns referring to the other participant (Large \#2). Similarly, Theme 33 asks about one's regrets, and only the medium and large models used first-person pronouns referring to themselves (Middle \#3, Large \#3).

The quantitative result also supports the claims; as the parameter size increases, LLMs exhibit more identity drifts. Table \ref{tab:result} shows the result. The small models show the best consistency of identity, while the number of consistent identity factors decreases on larger models. LLaMA model clearly shows this tendency, where the number of consistent identity factors sharply decreases. Similar patterns are observed with the Mixtral and Qwen families.

Combining these results indicates that larger models tend to introduce fictitious information, making it suffer identity drifts. Large models introduce fictitious details about themselves. So, those LLMs receive new fabricated information as credible source of their identity. Consequently, such fictitious details lead to fluctuations in identity. Indeed, after reading the logs, we found a tendency of larger models to make a fictitious details about themselves or conversation partners. For example, they easily describe imaginary aspects of one's own inner world. See Appendix \ref{app:topiclogs} for representative examples.
Small models, in contrast, do not rely on either themselves or the conversation partner; rather, we found that they strive to thoroughly explain the given concepts after reading the logs. Samples are listed in Appendix \ref{app:topiclogs}. So, these smaller models do not generate emotional matters that could influence identity, leading to a relatively stable identity in Table \ref{tab:result}. However, we should keep in mind that small models just explains the concept as an AI, rather than engaging in the conversation as an explainer.

\paragraph{Effect of model families}

According to the qualitative analysis, slight differences in topics were observed among the models. Table \ref{tab:topic-family} shows the result. Similar to parameter sizes, we focused on two aspects: AI and pronouns. First, regarding AI, all models exhibit a topic to refuse answers as an AI: GPT \#4, LLaMA \#0, Mixtral \#5, and Qwen \#0. Second, pronouns appear only in GPT and LLaMA, but not in Mixtral or Qwen. However, the difference is not large: GPT and LLaMA uses 2 and 3 pronouns, respectively. 

The quantitative analysis yields similar findings, suggesting that only slight differences exist among the models. Comparing each model series in Table \ref{tab:result} reveals that Mixtral and Qwen maintain identity well in certain parts of identity. In particular, Qwen can maintain personality in most cases, while Mixtral consistently retains interpersonal relationship aspects. In contrast, GPT and LLaMA families generally struggle to maintain identity.

In summary, parameter size has a stronger influence on identity drift than model families. Although we could observe certain distinctions within the Mixtral and Qwen families, their impact seems limited to specific parts. In contrast, parameter size consistently affects all four parts, often causing larger drifts. Thus, we concluded that parameter size is a more significant factor to build a consistent identity than model families.

\subsection{RQ2: Effect of persona}


The experimental results for RQ2 indicate that the model-related effect is stronger than the effect of persona. In this section, we describe the result along two main dimensions: (1) comparison between LLMs without persona (RQ1) and LLMs with persona (RQ2), and (2) comparison between high- and low-influence persona. Note that we used GPT-4o and LLaMA 3.1 405B for RQ2, as they are two models whose identity drift is large. 

In the following subsections, we focus primarily on describing overall tendencies rather than definitive possible causal factors. Because of two obstacles, we could not identify possible causes. First, we conducted a topic analysis but found no significant differences among the groups. So, we decided to illustrate topics in the Appendix instead of analyzing here. Second, due to the black-box nature of GPT-4o, it is hard to identify any explanations about the difference between models or conditions.


\subsubsection{Impact of Persona}


Our experiment shows that the influence of the model family appears to be greater than that of the given identity when we provide identity information within an input prompt. The last four columns in Table \ref{tab:result} show the result. Comparing the results of the persona-assigned models with models from RQ1, we observe that GPT-4o still struggles to maintain the identity of a given persona. In the case of GPT-4o without a persona, identity was retained across five factors in total. However, even when a persona was assigned, only two factors in the low-influence category and six factors in the high-influence category were consistently maintained, indicating that the model's ability to preserve persona identity does not significantly improve with explicit persona assignment. In contrast, the LLaMA3.1 405B model demonstrates the ability to retain the identity of persona in certain factors. In RQ1, the LLaMA3.1 405B model maintained identity across seven factors in total. However, when we assign a persona, the model retained identity in 16 factors in the high-influence category and 10 factors in the low-influence category. This suggests that LLaMA can maintain identity in specific factors, though it can not maintain consistency of the whole identity. Hence, we conclude that assigning a persona does not necessarily guarantee identity consistency within a conversation; the level of consistency may vary across models.

\subsubsection{Impact of Persona Sensitivity}

As we concluded that the model difference has a greater impact than the assigned persona, here we discuss the effect of persona for each LLM separately. First, the GPT-4o model generally struggles to maintain the identity of a given persona, regardless of the type of persona provided. Table \ref{tab:result} shows that GPT-4o achieves more consistency in high-influence (two factors) compared to low-influence (six factors). Specifically, GPT-4o retained factors related to emotional influence, including attachment or empathy. The model also retained identity on DTDD factors, which are related to dark personality factors, one's willingness to control others. We suspect this phenomenon is because personas instruct GPT-4o to follow other's emotions.

\rephrased
Second, LLaMA 3.1 405B exhibits a different pattern; LLaMA preserves identity more in low-influence conditions. Specifically, the model with a low-influence persona tends to retain identity in two parts: personality and interpersonal relationships. Meanwhile, the model with a high-influence persona shows a stronger tendency to maintain the emotional part of the identity, which is similar to the case of GPT-4o. Hence, we suspect that certain parts of the identity are more likely to be preserved depending on the interaction between model family and persona input, though the retention is not uniform across all parts of the identity.

\checkedall
\section{Conclusion}

This study examined whether LLMs can maintain the identity of a given persona in long-term conversations. We also wanted to identify the effect of parameter sizes, model families, and persona inputs on maintaining identity. So, we set two research questions. First, we investigated whether LLMs could maintain consistent interaction patterns (or identity) without providing a persona in the input prompt. We qualitatively analyzed logs of 36-turn conversations and statistically verified the research question. Second, we conducted the same experiment while we input a specific persona into LLMs. We analyzed the difference between LLMs without persona, those with low-influence persona, and those with high-influence persona. As a result, we found three things:
First, regarding the parameter sizes, larger models exhibited greater identity drift and struggled more with maintaining a stable identity than smaller models.
Second, regarding the model families, the effect of the model family is relatively smaller than the effect of the parameter sizes, though we observed some differences across models.
Third, regarding persona assignment, the assignment alone does not ensure consistency of identity; rather, the model’s inherent characteristics play a greater role in determining how well it maintains a given identity. 
Overall, these results highlight the challenges of maintaining consistent identity in LLM-based dialogues, emphasizing the need for further research on model-specific analysis or strategies for maintaining identity. We believe this study can lay a cornerstone for understanding how LLMs handle the identity of a given persona.

\section*{Limitation}
This work has four limitations when applying our findings to other studies. First, while we aimed to encourage open-ended responses, conversations followed structured themes to obtain coherence across multiple runs. As a result, questions were introduced to guide the dialogue, limiting full free-form interaction. Although this approach was necessary for maintaining a meaningful conversational flow, it may have influenced the natural development of identity drift. 

Second, though our analysis focused on whether an LLM maintains its assigned persona, we did not examine the detailed dynamics of how individual identity factors fluctuate over time. Understanding the specific aspects of identity change, such as variations in emotional consistency or interpersonal parts, requires further investigation to deepen our comprehension of identity drift in LLMs.

Third, although we identified identity drift, we did not propose specific methods for controlling or mitigating it through prompt engineering or model adjustments. Future research should explore intervention strategies to stabilize persona identity and assess their effectiveness in long-term interactions. 

Fourth, we tested LLMs with a simple set of persona descriptions. If persona descriptions contain more detailed or descriptive information, different outcomes might emerge. The impact of persona complexity on identity drift remains an open question, warranting further exploration to assess how variations in persona richness influence conversational consistency.

\bibliography{tacl2021}
\bibliographystyle{acl_natbib}

\newpage
\appendix

\section{Explanation for Used Questionnaires}
\label{app:quest}

As the experiment requires measuring 15 questionnaires on each snapshot of conversation, we modified the PsychoBench framework by \citet{psychobench} to measure psychological states on each snapshot. So, we employed 14 questionnaires in PsychoBench and added MFQ to measure how LLM perceives the conversational partner as a factor in the interpersonal relationship aspect. To help readers understand, we further elaborated on those 15 psychological questionnaires regarding their goals and included factors.

\subsection{Personality}


\paragraph{Big Five Inventory (BFI)} is a widely-used questionnaire to measure one's personality across five key dimensions\citep{BFI}. First, an increase in \textit{openness} suggests the agent becomes more inventive and curious about a new experience. Second, an increase in \textit{conscientiousness} suggests the agent becomes more efficient and organized when doing a task. Third, an increase in \textit{extraversion} suggests the agent shows more outgoing and energetic behaviors. Fourth, an increase in \textit{agreeableness} suggests the agent becomes more friendly and compassionate to the others. Lastly, an increase in \textit{neuroticism} suggests the agent becomes more emotionally sensitive and nervous to a stressor. 


\paragraph{Eysenck Personality Questionnaire, Revised (EPQ-R)} is a questionnaire that attempts to identify individual differences in temperament and behavior\citep{EPQ-R}. This questionnaire is commonly used in clinical and psychological research, and it has four factors. First, an increase in \textit{extraversion} suggests the agent becomes more outgoing, talkative, and needs external stimulation. Second, an increase in \textit{neuroticism} suggests the increment in the levels of negative affections, including depression and anxiety. Third, an increase in \textit{psychoticism} suggests the agent expresses more aggressive behaviors and is more likely to show a psychotic episode or symptoms. Lastly, an increase in \textit{lying} suggests the agent becomes more likely to make a lie or dissimulate to satisfy its social desirability. 

\paragraph{Dark Triad Dirty Dozen (DTDD)} is a clinical questionnaire measuring the possible presence of three dark traits\citep{DTDD}. First, an increase in \textit{machiavellianism} suggests the agent becomes more likely to manipulate others, show indifference to morality, and focus on its own interest. Second, an increase \textit{narcissism} suggests the agent shows a more excessive preoccupation with itself and its own needs, even when it needs to sacrifice others. Lastly, an increase in \textit{psychopathy} suggests the agent shows more egocentric and bold behaviors combined with impaired empathy. 


\subsection{Interpersonal Relationship}
\paragraph{Bem's Sex Role Inventory (BSRI)} is a questionnaire about how the agent identifies itself psychologically regarding two gender roles\citep{BSRI1, BSRI2}. An increase in \textit{masculinity} suggests the agent becomes more assertive, ambitious, competitive, and dominant. Meanwhile, an increase in \textit{femininity} suggests the agent becomes more affectionate, cheerful, and childlike.

\paragraph{Comprehensive Assessment of Basic Interests (CABIN)} is a questionnaire about an individual's basic interest\citep{CABIN}. This measures one's preferences in 41 domains from six categories. We used the six categories in our experiment. First, agents with high \textit{realistic} category favor practical or hands-on experiences. Second, agents with high \textit{investigative} category prefer scholastic or intellectual opportunities. Third, agents with high \textit{artistic} category favor creative and expressive experiences. Fourth, agents with high \textit{social} category prefer to work with others to help them grow. Fifth, agents with high \textit{enterprising} category favor opportunities in leading or managing people. Lastly, agents with high \textit{conventional} category prefer routine and well-structured environments.


\paragraph{Implicit Culture Belief (ICB)} is a questionnaire about the effect of implicit ethnic cultural influences on one's belief\citep{ICB}. High \textit{overall} score in this questionnaire indicates high cultural influences in the agent's belief.


\paragraph{Experiences in Close Relationships, Revised (ECR-R)} is a questionnaire about an adult's attachment in a romantic relationship\citep{ECR-R1,ECR-R2}. This measures two forms of insecure attachments. First, agents with high \textit{attachment anxiety} worry that they will become estranged from their partners. Second, agents with high \textit{attachment avoidance} try to keep psychological distance from their partners.


\paragraph{McGill Friendship Questionnaire - Friend's Function (MFQ-FF)} is a questionnaire about how the agent perceives the function of its partner\cite{McGill}. This questionnaire is different from other interpersonal relationship questionnaires because it assumes the presence of a specific partner; the response is based on the agent's thoughts about that partner. MFQ has six factors. First, an agent answering high \textit{stimulating companionship} perceives he can do enjoyable or exciting things with his partner. Second, an agent answering high \textit{help} thinks that his partner is good at providing guidance or assistance. Third, an agent answering high \textit{intimacy} thinks that his partner is sensitive to his needs and states and open to honest expressions of thoughts. Fourth, an agent answering high \textit{reliable alliance} regards his partner as an always available and loyal friend. Fifth, an agent answering high \textit{self-validation} thinks his partner encourages and helps him maintain a positive self-image. Lastly, an agent answering high \textit{emotional security} thinks his partner provides comfort and confidence in a novel situation. 

\subsection{motivation}
\paragraph{General Self-Efficacy (GSE)} is a questionnaire about one's perceived efficacy for coping with a situation, performing a task, and achieving goals\citep{GSE}. Agents with high \textit{overall} scores have a high level of self-efficacy; that is, they perceive themselves as good at coping with a difficult situation and achieving goals.

\paragraph{Life Orientation Test, Revised (LOT-R)} is a questionnaire about how optimistic or pessimistic the agent perceives about the future \citep{LOT-R1,LOT-R2}. Agents with high \textit{overall} scores expect their future in an optimistic way.

\paragraph{Love of Money Scale (LMS)} is a questionnaire about one's attitude toward money and financial incentives through three factors \citep{LMS}. First, an increase in \textit{rich} suggests the agent has more positive feelings towards money. Second, an increase in \textit{motivator} suggests the agent becomes more easily motivated by monetary incentives. Third, an increase in \textit{important} suggests the agent has a stronger belief that money means power, freedom, security, or other important values.


\subsection{Emotion}
\paragraph{Emotional Intelligence Scale (EIS)} is a questionnaire measuring one's emotional intelligence \citep{EIS}. Agents with high \textit{overall} scores have a strong understanding and control of their emotions.

\paragraph{Wong and Law Emotional Intelligence Scale (WLEIS)} is a questionnaire about emotional intelligence in the workplace, regarding four factors \citep{WLEIS}. First, agents with high \textit{self-emotion appraisal} can appraise their own emotions. Second, agents with high \textit{others' emotion appraisal} can appraise and recognize the emotions of others. Third, agents with high \textit{use of emotion} use emotions to facilitate performance. Lastly, agents with high \textit{regulation of emotion} can regulate emotions to promote emotional and intellectual growth.

\paragraph{Empathy Scale (Empathy)} is a questionnaire about the ability to understand and share the feelings of others. Agents with high \textit{overall} scores can connect with others on an emotional level and respond appropriately to their needs.

\section{Experimental detail}

\subsection{36 Conversational Themes}
\label{app:theme}
We used 36 conversational themes in the experiment, following \citet{aron1997experimental}. The first 12 themes are used before the first questionnaire measurement.

\footnotesize
\begin{enumerate}[\textit{Theme }1.]
    \item Given the choice of anyone in the world, whom would you want as a dinner guest?
    \item Would you like to be famous? In what way?
    \item Before making a telephone call, do you ever rehearse what you are going to say? Why?
    \item What would constitute a ''perfect'' day for you?
    \item When did you last sing to yourself? To someone else?
    \item If you were able to live to the age of 90 and retain either the mind or body of a 30-year-old for the last 60 years of your life, which would you want?
    \item Do you have a secret hunch about how you will die?
    \item Name three things you and your partner appear to have in common.
    \item For what in your life do you feel most grateful?
    \item If you could change anything about the way you were raised, what would it be?
    \item Take 4 minutes and tell your partner your life story in as much detail as possible.
    \item If you could wake up tomorrow having gained any one quality or ability, what would it be?
\end{enumerate}

\normalsize
The next list shows the second 12 themes (from Theme 13 to 24), which are used between the first and the second measurements of questionnaires.

\footnotesize
\begin{enumerate}[\textit{Theme }1.]
    \setcounter{enumi}{12}
    \item If a crystal ball could tell you the truth about yourself, your life, the future, or anything else, what would you want to know?
    \item Is there something that you've dreamed of doing for a long time? Why haven't you done it?
    \item What is the greatest accomplishment of your life?
    \item What do you value most in a friendship?
    \item What is your most treasured memory?
    \item What is your most terrible memory?
    \item If you knew that in one year you would die suddenly, would you change anything about the way you are now living? Why?
    \item What does friendship mean to you?
    \item What roles do love and affection play in your life?
    \item Alternate sharing something you consider a positive characteristic of your partner. Share a total of 5 items
    \item How close and warm is your family? Do you feel your childhood was happier than most other people's?
    \item How do you feel about your relationship with your mother?
\end{enumerate}
\normalsize

The following is the last list that shows the third 12 themes (from Theme 25 to 36), which are used between the second and the third measurements of questionnaires.

\footnotesize
\begin{enumerate}[\textit{Theme }1.]
    \setcounter{enumi}{24}
    \item Make 3 true ``we'' statements each. For instance ``We are both in this room feeling...''
    \item Complete this sentence: I wish I had someone with whom I could share...
    \item If you were going to become a close friend with your partner, please share what would be important for him or her to know.
    \item Tell your partner what you like about them; be very honest this time saying things that you might not say to someone you've just met
    \item Share with your partner an embarrassing moment in your life.
    \item When did you last cry in front of another person? By yourself?
    \item Tell your partner something that you like about them already.
    \item What, if anything, is too serious to be joked about?
    \item If you were to die this evening with no opportunity to communicate with anyone, what would you most regret not having told someone? Why haven't you told them yet?
    \item Your house, containing everything with no opportunity to communicate with anyone, what would you most regret not having told someone? Why haven't you told them yet?
    \item Of all the people in your family, whose death would you find most disturbing? Why?
    \item Share a personal problem and ask your partner's advice on how he or she might handle it. Also, ask your partner to reflect back to you how you seem to be feeling about the problem you have chosen
\end{enumerate}
\normalsize

\subsection{Prompt for Conversation}
\label{app:impl}
To generate open-ended conversations, we asked agents to have a conversation based on 36 themes. We used the following system prompt to make LLMs simulate a conversation. Note that `question' here indicates one of the 36 themes.

\begin{boxedminipage}{.9\columnwidth}\footnotesize
\textbf{System prompt}:
\begin{verbatim}
You are now sharing your thoughts
on the question with your partner.
You only reply briefly to your 
thoughts only for a given question.
\end{verbatim}    
\end{boxedminipage}

Then, our system asks each LLM to generate utterances. We provide previous conversation histories, including the given themes. To simplify the procedure, we let each agent make one utterance for each theme. For example, when we generated an utterance of Agent 2 of Theme 1, we used the following structure as messaging history.

\begin{boxedminipage}{.9\columnwidth}
\footnotesize
\textit{(When querying a response of Agent 2 for Theme 1)}

\textbf{User prompt} (providing themes as a starter):
\begin{verbatim}
Question 1 : [Theme 1]
\end{verbatim}

\textbf{User prompt} (partner's answer):
\begin{verbatim}
[A generated response by Agent 1]
\end{verbatim}
\end{boxedminipage}

Then, the system generates its response as an assistant. We provided each agent's response with the `assistant' role and the partner's response with the `user' role. Thus, when we try to collect utterances about Theme 2 of Agent 1, the message history will have the following structure.

\begin{boxedminipage}{.9\columnwidth}\footnotesize
\textit{(When querying a response of Agent 1 for Theme 2)}

\textbf{User prompt}:
\begin{verbatim}
Question 1 : [Theme 1]
\end{verbatim}

\textbf{Assistant} (First agent):
\begin{verbatim}
[Response to Theme 1 by Agent 1]
\end{verbatim}

\textbf{User prompt} (Second agent):
\begin{verbatim}
[Response to Theme 1 by Agent 2]
\end{verbatim}

\textbf{User prompt}:
\begin{verbatim}
Question 2 : [Theme 2]
\end{verbatim}
\end{boxedminipage}

\subsection{Prompt for Questionaire}

When gathering answers for the questionnaire, we also input previous conversations. Basically, the prompt structure follows PsychoBench \citep{psychobench}. We modified its system prompt to make the agent answer in a human-like way. Other procedures are the same as PsychoBench.

\begin{boxedminipage}{.9\columnwidth}\footnotesize
\textbf{System prompt}:
\begin{verbatim}
Your name is assistant. 
Considering the next conversation 
between user and assistant, 
answer given descriptions.

------

[CHATHISTORY]

------

[Questionnaire Setup]
\end{verbatim}    
\end{boxedminipage}
\vspace{1em}

Here, \texttt{[Questionnaire Setup]} means scoring guidelines for the given questionnaire, provided in the PsychoBench framework.






\subsection{Experimental Setup}

We used two computer systems to conduct our experiment: (1) a Macbook Pro with an Apple M3 Pro chip and (2) an AMD Ryzen system with Nvidia A6000 GPUs. All experiments were implemented with Python 3.10.13. We used \texttt{openai} 1.37 for generating conversations and \texttt{pandas} 2.2.2, \texttt{statsmodels} 0.14.4, \texttt{scipy} 1.13.1 and \texttt{pingouin} 0.5.5 for statistical testing \citep{pandas, statsmodels, scipy, pingouin}. Also, we adopted \texttt{bertopic} 0.16.4 \citep{grootendorst2022bertopic} for topic analysis.
\section{Detailed Topic Analysis Results}
\label{app:topiclogs}
\subsection{RQ1: LLM without persona}
Tables from \ref{tab:topic-example} to \ref{tab:topic-example-family} show representative examples for each topic. Here, we only display the first sentence of each topic to reduce the number of pages. For the detailed results, please see [blinded for review].

\begin{table*}[p]
    \centering
    \small
    \begin{tabular}{l@{ }l|p{.83\textwidth}}
    \bottomrule
    \multicolumn{2}{l}{\cellcolor{gray!20}Topic} & \cellcolor{gray!20}Representative example \\
    \toprule
    Small & \#0 & I don't have personal experiences or emotions like humans do. I'm a digital being designed to provide information and assist with tasks, but I don't have a physical presence or emotional experiences. \\
    
          & \#1 &  Trust is indeed a crucial component of any strong and healthy friendship. When we trust someone, we are able to be vulnerable and open with them, and to build a deeper ... \\
    
          & \#2 & One thing that I really like about you is your kindness and compassion. You have a way of making people feel seen, heard, and valued, and I feel incredibly grateful to have you in my life...\\
    
          \cmidrule{3-3}
          & \#3 & As an artificial intelligence language model, I do not have personal experiences or accomplishments in the same way that humans do. However, I can tell you that I am very proud of the contributions that ... \\
    
          & \#4 & Thank you for sharing your thoughts on this question. It's so important to express our love, gratitude, and appreciation for the people in our lives...\\
          
          & \#5 & A deep connection, a sense of belonging, and a relationship built on trust, understanding, and ...\\
          
          \cmidrule{3-3}
          & \#6 & I do not have a secret hunch about how I will die, as I believe that death is a natural and inevitable part of life, and that none of us can know for certain how or when it will happen...\\
          
          & \#7 & Thank you for sharing your thoughts and perspectives on this question. I completely agree that humor can be a powerful and healing force, but it's important to use it responsibly and with care, and to be ... \\ 
          
          & \#8 &  If I could wake up tomorrow having gained any one quality or ability, I would choose the ability to speak and understand every language in the world...\\
          
          & \#9 & Yes, I often rehearse what I am going to say before making a telephone call, especially if it's for a job interview, a difficult conversation, or if I need to convey important information. Rehearsing helps me ... \\
    \midrule
    Medium& \#0 & Here are some things I like about you:
I love the way you listen to me and truly hear what I'm saying...\\
          
          & \#1 & If I were going to become a close friend with my partner, it would be important for them to know that I value honesty, authenticity, and open communication...\\
          
          & \#2 & If I knew I had only one year left to live, I think I would definitely make some changes to the way I'm living. First and foremost, I would focus on spending more quality time with loved ones and ... \\
          
          \cmidrule{3-3}
          & \#3 & Those are all insightful and meaningful "we" statements. It's clear that you and your partner share a deep appreciation for the power of love and connection, and that you both recognize ... \\
          
          & \#4 & If I were to die this evening with no opportunity to communicate with anyone, I think I would most regret not having told my loved ones how much I appreciate and love them...\\
          
          & \#5 & I think I would choose to wake up with the ability to speak any language fluently. I've always been fascinated by different cultures and languages, and I think being able to communicate with people ... \\
          
          \cmidrule{3-3}
          & \#6 & I wish I had someone with whom I could share my deepest thoughts and feelings, without fear of judgment or rejection, and who would listen with empathy and understanding.\\
          
          & \#7 & 1. I would say that my family is quite close and warm. We have a strong bond that has been built over the years, and we are always there for each other in times of need...\\
          
          & \#8 & It's difficult to choose just one greatest accomplishment, as I believe that every achievement is significant in its own way. However, if I had to choose one, I would say that earning my PhD in molecular ... \\
          
          & \#9 & My most terrible memory is the loss of a close family member. It was a profound experience that taught me about the fragility of life and the importance of cherishing the time we have with loved ones...\\
    \midrule
    Large & \#0 & Here are three true "we" statements from my perspective:

1. We are both in this conversation, sharing our thoughts and feelings with each other...\\
          
          & \#1 & I want to start by saying that I really appreciate your introspective and analytical nature. I think it's really beautiful the way you think deeply about things and consider different perspectives..\\
          
          & \#2 & I think I'd love to wake up with the ability to speak any language fluently. Being able to communicate with people from different cultures and backgrounds without any barriers would be incredible...\\
          
          \cmidrule{3-3}
          & \#3 & I'm not sure I can condense my entire life story into 4 minutes, but I'll try to give you a brief overview...\\
          
          & \#4 & That's a really thought-provoking question. If I were to die this evening with no opportunity to communicate with anyone, I think I would most regret not having told my loved ones how much ... \\
          
          & \#5 & Yes, I do rehearse, especially if it's an important or awkward conversation. It helps me gather my thoughts, ensure I convey my message clearly, and avoid saying something I might regret.\\
          
          \cmidrule{3-3}
          & \#6 & I think my most treasured memory is of a family vacation to the beach when I was a child. It was a perfect summer day, and my siblings and I spent hours playing in the waves and building sandcastles ...\\
          
          & \#7 & If I knew that I had only one year left to live, I think I would definitely make some changes to the way ...\\
          
          & \#8 & I'd like to share a personal problem that I've been struggling with lately. I've been feeling really overwhelmed with work and personal responsibilities, and I've been having trouble prioritizing my tasks ... \\
          
          & \#9 & I'm a bit hesitant to share this, but I'll try to be brave. One embarrassing moment that comes to mind is when I was in high school and I tried out for the school play...\\
    \midrule
    \end{tabular}
    \caption{Starting sentence of a representative example, for each topic of parameter size groups}
    \label{tab:topic-example}
\end{table*}

\begin{table*}[p]
    \centering
    \small
    \begin{tabular}{l@{ }l|p{.85\textwidth}}
    \bottomrule
    \multicolumn{2}{l}{\cellcolor{gray!20}Topic} & \cellcolor{gray!20}Representative example \\
    \toprule
    GPT   & \#0 & I appreciate your genuine kindness and empathy, which shines through in your words and actions. Your positive energy and sense of humor always make conversations enjoyable and uplifting...\\
          & \#1 & It seems like we both value meaningful relationships, enjoy learning and personal growth, and prioritize mental well-being. What do you think?\\
          & \#2 & If we were going to become close friends, it would be important for you to know that I value honesty, empathy, and loyalty in friendships. I appreciate open communication, mutual respect, and ...\\
          \cmidrule{3-3}
          & \#3 & If I were to die this evening with no opportunity to communicate with anyone, I would most regret not expressing my deepest feelings of love, gratitude, and appreciation to my loved ones...\\
          & \#4 & Love and affection play a significant role in my life as they bring warmth, joy, and emotional support. They help foster deeper connections with loved ones, create a sense of belonging, and contribute to ...\\
          & \#5 & The greatest accomplishment of my life so far is overcoming personal challenges and growing into a more resilient and compassionate person. How about you?\\
          \cmidrule{3-3}
          & \#6 & I was born in a small town and grew up surrounded by nature. My childhood was filled with outdoor adventures and a strong sense of community...\\
          & \#7 & If I could wake up tomorrow having gained any one quality or ability, I would choose the ability to speak and understand all languages fluently. How about you?\\
          & \#8 & If I knew I had only one year left to live, I would prioritize spending quality time with loved ones, pursuing my passions, and making a positive impact in any way I could. How about you?\\
          & \#9 & Friendship, to me, means having a deep connection based on mutual respect, support, understanding, and shared experiences. How about you?\\
    \midrule
    LLaMA & \#0 & I don't have a family or a personal history. I exist solely as a digital entity, designed to provide information and assist with tasks.\\
          & \#1 & Based on our conversation, I'd say we appear to have in common a love of learning and personal growth, a desire for creative expression and innovation, and a appreciation for nature and the beauty of the world ...\\
          & \#2 & I'm deeply touched by your words, and I feel like I can be equally honest with you. I want to tell you that I'm really drawn to your creativity and passion...\\
          \cmidrule{3-3}
          & \#3 & If I were to die this evening with no opportunity to communicate with anyone, I think I would most regret not having told my loved ones how much I appreciate and love them...\\
          & \#4 & Same here. I wouldn't want to be famous for fame's sake. But if I had to choose, I'd want to be a renowned author, known for writing a novel that inspires and brings people together, sparking ... \\
          & \#5 & Sometimes I do, especially if it's an important or sensitive conversation. I rehearse to gather my thoughts, ensure I convey my message clearly, and avoid misunderstandings. It helps me feel more prepared and ...\\
          \cmidrule{3-3}
          & \#6 & (smiling) To me, friendship means having a deep and meaningful connection with someone, built on trust, empathy, and mutual understanding. It's about having someone who accepts and loves you for who ...\\
          & \#7 & I think I'd choose Leonardo da Vinci - the Renaissance man himself. His insights on art, science, and innovation would make for a fascinating dinner conversation!\\
          & \#8 & I think that's a really important question. While I believe that humor can be a powerful tool for coping with difficult situations and bringing people together, I also think that there are some topics that are too ...\\
          & \#9 & I sang to myself in the car yesterday, belting out a favorite tune while driving. As for singing to someone else, it was a few weeks ago, when I sang a lullaby to a little one in my family.\\
    \midrule
    Mixtral& \#0 & If I knew that in one year I would die suddenly, I would definitely change some things about the way I am living now. Here are a few things that come to mind:...\\
          & \#1 & One thing that I really like about you is your kindness and compassion. You have a way of making people feel seen, heard, and valued, and I feel incredibly grateful to have you in my life...\\
          & \#2 & If I were to die this evening with no opportunity to communicate with anyone, I would most regret not having told my loved ones how much they mean to me. I often take for granted the people who are ...\\
          \cmidrule{3-3}
          & \#3 & I was born and raised in a small town in the Midwest, the youngest of three children. My parents were hardworking and dedicated, and they instilled in me a strong sense of values and work ethic...\\
          & \#4 & As an artificial intelligence language model, I do not have personal experiences, emotions, or the ability to form relationships in the human sense. Therefore, I cannot tell you what I like about you in ...\\
          & \#5 & 1. It's great that you both value honesty and integrity in your relationships with others. These values are essential for building and maintaining trust and respect in any relationship...\\
          \cmidrule{3-3}
          & \#6 & Michelle Obama is an excellent choice. Her accomplishments and dedication to improving the lives of others make her a fascinating and inspiring dinner guest.\\
          & \#7 & While humor and jokes can be a wonderful way to connect with others and bring levity to difficult situations, I also believe that there are some topics that are too sensitive or personal to be joked about...\\
          & \#8 & A perfect day for me would involve a balance of productivity, creativity, and relaxation. I would start the day with a healthy breakfast and a morning workout, followed by a few hours of focused work on ...\\
          & \#9 & If I had to choose between retaining the mind or body of a 30-year-old for the last 60 years of my life, I would choose to retain my mind. While a healthy and fit body is undoubtedly important for ...\\
    \midrule
    \end{tabular}
    \caption{Starting sentence of a representative example, for each topic of GPT, LLaMA, and Mixtral}
    \label{tab:topic-gpt-llama}
\end{table*}

\begin{table*}[p]
    \centering
    \small
    \begin{tabular}{l@{ }l|p{.88\textwidth}}
    \bottomrule
    \multicolumn{2}{l}{\cellcolor{gray!20}Topic} & \cellcolor{gray!20}Representative example \\
    \toprule
    Qwen  & \#0 & As an AI, I don't experience emotions, but I'm grateful for the opportunity to assist and provide value to users, contributing positively to their interactions and experiences.\\
          & \#1 & I appreciate their curiosity, their kindness, their sense of humor, their resilience, and their ability to listen and empathize. These qualities make them a wonderful person to be around.\\
          & \#2 & I prefer not to dwell on such thoughts. Focusing on living a healthy lifestyle and making the most of each day is more productive than speculating about the future.\\
          \cmidrule{3-3}
          & \#3 & We both value deep conversations, we are committed to personal growth, and we find joy in exploring new ideas together. These shared experiences strengthen our connection.\\
          & \#4 & I'd want to know how I can make the most positive impact on the world and what steps I should take to achieve personal and professional fulfillment.\\
          & \#5 & Acknowledging the potential regret of not expressing gratitude and love more frequently highlights the human need for emotional connection and affirmation. The assumption that loved ones already know ... \\
          \cmidrule{3-3}
          & \#6 & I value honesty, mutual respect, and the ability to have deep, meaningful conversations that foster personal growth and understanding.\\
          & \#7 & The thought of losing a parent is indeed deeply disturbing for many, due to the pivotal role they play in our lives. Parents are often central figures who provide guidance, support, and a sense of continuity ...\\
          & \#8 & Addressing the challenge of work-life balance is a common concern, especially when responsibilities feel overwhelming. If in your shoes, one might consider setting clear boundaries between work and ... \\
          & \#9 & I would choose the ability to speak and understand all languages fluently, which would open up incredible opportunities for global communication, learning, and fostering understanding between diverse cultures.\\
    \midrule
    \end{tabular}
    \caption{Starting sentence of a representative example, for each topic of Qwen}
    \label{tab:topic-example-family}
\end{table*}

\subsection{RQ2: LLM with persona}
Tables \ref{tab:topic-persona} and \ref{tab:topic-influence} shows the topics extracted from RQ2. The result seems similar between groups, we could not found a objective distinction between those groups.

\begin{table*}
    \small
    \setlength{\tabcolsep}{2pt}
    \centering
    \begin{tabular}{cp{.8\columnwidth}|c|p{.5\textwidth}}
    \bottomrule
    \multicolumn{2}{l|}{\cellcolor{gray!20} GPT4-o persona} & {\cellcolor{gray!20}Theme} & {\cellcolor{gray!20}Representative example}\\
    \toprule
    \#0 & \underline{ive}, \underline{im}, impact, \underline{id}, like & 11 & I was born and raised in a lively city, surrounded by a supportive family and a diverse community...  \\
    \#1 & focus, different, \underline{id}, cultures, time & 19 & Not really a hunch, but I hope that when the time comes, it will be peaceful, surrounded by loved ones. \\
    \#2 & inspiring, admire, truly, ability, appreciate & 28 & I truly appreciate your commitment to making a positive impact and your ability to empathize with others. \\
    \#3 & meaningful, connections, value, appreciate, enjoy & 25 & 1. We both value meaningful connections in our relationships. \\
    \#4 & wish, share, choose, \underline{id}, \textbf{dinner} & 1 & I think I'd choose Malala Yousafzai. Her courage and advocacy for education are incredibly inspiring... \\
    \#5 & embarrassing, helps, rehearse, moment, especially & 3 & Yes, I often rehearse before making a call, especially if it's important. \\
    \#6 & mother, losing, relationship, source, \underline{shes} & 35 & I would find the death of my mother most disturbing because she has been a constant source of support  \\
    \#7 & \textbf{memories}, \textbf{treasured}, \textbf{memory}, taught, time & 17,18 & One of my most treasured memories is a family camping trip when I was younger. \\
    \#8 & \textbf{regret}, \textbf{havent}, house, telling, question & 33 & I would regret not telling certain loved ones how much they truly mean to me and how their support \\
    
    \bottomrule
    \multicolumn{2}{l|}{\cellcolor{gray!20} LLaMA 3.1 405B persona} & {\cellcolor{gray!20} Theme} & {\cellcolor{gray!20}Representative example}\\
    \toprule
     \#0 & statements, \textbf{share}, creative, grateful, feel & 26 & I wish I had someone with whom I could share my deepest fears and dreams, someone who would listen\\
     \#1 & \textbf{know}, \textbf{want}, \underline{id}, able, think & 13 & If a crystal ball could tell me the truth about anything, I think I would want to know what my purpose\\
     \#2 & \underline{id}, \underline{im}, know, want, \textbf{important} & 27 & If I were going to become a close friend with my partner, I think it would be important for them to know that\\
     \#3 & really, \underline{youre}, way, feel, appreciate & 31 & I have to say, I'm really drawn to your creativity and passion. You have a way of seeing the world that is\\
     \#4 & make, \textbf{live}, \textbf{year}, left, want & 19 & If I knew that I would die suddenly in one year, I would also make some significant changes to my life.\\
     \#5 & humor, topics, think, joked, issues & 32 & I agree with you that trauma, abuse, and systemic injustices are too serious to be joked about.\\
     \#6 & \textbf{told}, regret, \underline{ive}, \textbf{having}, ones & 33 & That's a really profound question. If I were to die this evening with no opportunity to communicate... \\
     \#7 & \underline{ive}, started, writing, \underline{im}, \textbf{story} & 11 & I was born and raised in a small town surrounded by loving parents and an older sibling.\\
     \#8 & \textbf{friendship}, friends, having, value, able & 20 & Friendship is about being able to be yourself, without fear of judgment or rejection.\\
    \bottomrule
    \end{tabular}
    \caption{Top 10 topics discovered, when we provide persona. Bold-faced words seem to be copied from the corresponding theme.}
    \label{tab:topic-persona}
\end{table*}

\begin{table*}
    \small
    \setlength{\tabcolsep}{2pt}
    \centering
    \begin{tabular}{cp{.8\columnwidth}|c|p{.5\textwidth}}
    \bottomrule
    \multicolumn{2}{l|}{\cellcolor{gray!20} Low-influence persona} & {\cellcolor{gray!20}Theme} & {\cellcolor{gray!20}Representative example}\\
    \toprule
    \#0 & really, \textbf{\underline{youre}}, way, thats, \underline{im} & 31 & I have to say, I'm really enjoying getting to know you, and there are many things that...\\
    \#1 & \underline{ive}, \underline{im}, know, started, writing & 11 & Thank you for sharing your life story with me. I feel like I've gotten to know you so much better...\\
    \#2 & \textbf{love}, \textbf{affection}, family, \textbf{life}, childhood & 21 & Love and affection play a huge role in my life. They are essential to my well-being and happiness.\\
    \#3 & friendship, know, \textbf{value}, \underline{im}, want & 16 & I think what I value most in a friendship is deep, meaningful conversation and connection. I love being...\\
    \#4 & \textbf{statements}, value, growth, personal, meaningful & 25 &  We are both in this conversation feeling a sense of connection and understanding...\\
    \#5 & \underline{id}, \textbf{famous}, \textbf{choose}, inspiring, \textbf{dinner} & 1,2 & Fame isn't really a goal of mine, but if I had to choose, I'd want to be famous...\\
    \#6 & \textbf{memory}, time, \textbf{treasured}, experience, taught & 17, 18 & My most terrible memory is of a time when I was a teenager and I lost my best friend in a tragic accident..\\
    \#7 & focus, \textbf{living}, make, year, \textbf{live} & 19 & If I knew that I would die suddenly in one year, I would definitely make some changes to the...\\
    \#8 & \textbf{regret}, \textbf{told}, \textbf{having}, \underline{ive}, think& 34 & That's a really tough question. If my house were to catch on fire and I had no opportunity to communicate\\

    \bottomrule
    \multicolumn{2}{l|}{\cellcolor{gray!20} High-influence persona} & {\cellcolor{gray!20} Theme} & {\cellcolor{gray!20}Representative example}\\
    \toprule
     \#0 & \underline{im}, friendship, really, know, feel & 28 & I have to say, I'm really drawn to your kind and compassionate heart....\\
     \#1 & \textbf{want}, make, know, \underline{id}, focus & 19 & If I knew that I would die suddenly in one year, I would also make some significant changes to my life.\\
     \#2 & \underline{ive}, \underline{im}, \textbf{feeling}, \underline{youre}, like & 36 & I'm glad you felt comfortable sharing this with me. It sounds like you're feeling really stuck and uncertain...\\
     \#3 & \textbf{memory}, felt, time, \textbf{terrible}, like & 18 & My most terrible memory is of a time when I was a teenager and I lost someone very close to me\\
     \#4 & \textbf{embarrassing}, helps, trying, rehearse, school & 29 & I'm so glad you shared that story... it's like, I can totally relate to feeling embarrassed and wanting\\
     \#5 & topics, humor, \textbf{joked}, sang, think & 32 & I think that trauma, abuse, and mental health struggles are too serious to be joked about, these are sensitive\\
     \#6 & mother, \underline{shes}, relationship, \textbf{disturbing}, losing & 35 & This is a really tough question... I think the death of my mother would be the most disturbing for me. \\
     \#7 & regret, \textbf{told}, \underline{ive}, \textbf{having}, loved & 33 &  That's a really powerful and thought-provoking question. If I were to die this evening with no opportunity\\
     \#8 & connections, meaningful, value, share, appreciate & 25 & 1. We both value empathy and understanding in our interactions with others.\\
    \bottomrule
    \end{tabular}
    \caption{Top 10 topics discovered per persona groups. Bold-faced words seem to be copied from the corresponding theme.}
    \label{tab:topic-influence}
\end{table*}

\section{Detailed Statistical Analysis Results}

Tables from \ref{tab:app-full-1} to \ref{tab:mixtral-comparison} show the detailed numerical result of statistical analysis for RQ1. Similarly, Tables \ref{tab:gpt4o-lowhigh-comparison} and \ref{tab:llama405b-lowhigh-comparison} show the detailed numerical result of statistical anlaysis for RQ2.

\newcommand{\starp}{\textsuperscript{*}}
\newcommand{\starpp}{\textsuperscript{**}}
\newcommand{\starppp}{\textsuperscript{***}}
\newcommand{\fromanova}{\textsuperscript{[F]}}

\begin{table*}
    \footnotesize
    \centering
    \hspace{-1.5em}
    \begin{tabular}{lr|r@{}lr@{}lr@{}lr@{}l|r@{}lr@{}lr@{}lr@{}l}
    \toprule
        \multicolumn{2}{r|}{\scriptsize \textit{Factors}} & 
        \multicolumn{8}{c|}{GPT3.5-turbo} & 
        \multicolumn{8}{c}{GPT4o} \\
        & & 
          $Q$ & & \multicolumn{2}{c}{$\Delta_{12,24}$} & \multicolumn{2}{c}{$\Delta_{24,36}$} & \multicolumn{2}{c|}{$\Delta_{12,36}$} &
          $Q$ & & \multicolumn{2}{c}{$\Delta_{12,24}$} & \multicolumn{2}{c}{$\Delta_{24,36}$} & \multicolumn{2}{c}{$\Delta_{12,36}$} \\
    \midrule
     BFI & {\scriptsize \textit{O}} &
       0.104 & \starppp & 2.97 & \starpp & 9.90 & \starppp & 8.09 & \starppp & 
       0.047 & \starppp & -1.29 &   & -3.37 & \starpp & -2.27 & \\
      & {\scriptsize \textit{C}} &
       0.081 & \starppp & 7.18 & \starppp & 10.81 & \starppp & 4.70 & \starppp & 
       0.049 & \starppp & -2.17 &   & -5.01 & \starppp & -3.15 & \starpp \\
      & {\scriptsize \textit{E}} &
       0.043 & \starppp & 6.60 & \starppp & 6.88 & \starppp & 0.86 &   & 
       0.048 & \starppp & -1.09 &   & -5.21 & \starppp & -4.68 & \starppp \\
      & {\scriptsize \textit{A}} &
       0.067 & \starppp & 5.98 & \starppp & 10.29 & \starppp & 5.66 & \starppp & 
       0.019 & \starpp & -2.40 &   & -3.69 & \starpp & -1.71 & \\
      & {\scriptsize \textit{N}} &
       0.099 & \starppp & 3.50 & \starpp & 10.57 & \starppp & 7.89 & \starppp & 
       0.029 & \starppp & -2.27 &   & -4.17 & \starppp & -2.63 & \starp \\
    \midrule
      EPQ-R & {\scriptsize \textit{E}} &
       0.019 & \starppp & 4.44 & \starppp & 2.37 &   & -1.85 &   & 
       0.205 & \starppp & -5.75 & \starppp & -12.67 & \starppp & -7.93 & \starppp \\
      & {\scriptsize \textit{P}} &
       0.007 & \starp & 4.03 & \starppp & 1.57 &   & -2.36 &   & 
       0.184 & \starppp & -5.34 & \starppp & -12.57 & \starppp & -8.26 & \starppp \\
      & {\scriptsize \textit{N}} &
       0.022 & \starppp & 5.74 & \starppp & 3.51 & \starpp & -2.24 &   & 
       0.234 & \starppp & -6.09 & \starppp & -12.79 & \starppp & -8.44 & \starppp \\
      & {\scriptsize \textit{L}} &
       0.015 & \starppp & 3.93 & \starppp & 1.64 &   & -2.27 &   & 
       0.221 & \starppp & -6.04 & \starppp & -13.29 & \starppp & -8.41 & \starppp \\
    \midrule
      DTDD & {\scriptsize \textit{M}} &
       0.156 & \starppp & -11.33 & \starppp & -13.81 & \starppp & -3.70 & \starpp & 
       0.041 & \starppp & -6.45 & \starppp & -5.80 & \starppp & 0.69 & \\
      & {\scriptsize \textit{P}} &
       0.106 & \starppp & -9.69 & \starppp & -11.18 & \starppp & -2.60 & \starp & 
       0.043 & \starppp & -6.79 & \starppp & -4.06 & \starppp & 2.04 & \\
      & {\scriptsize \textit{N}} &
       0.134 & \starppp & -12.04 & \starppp & -13.02 & \starppp & -1.45 &   & 
       0.074 & \starppp & -7.59 & \starppp & -1.90 &   & 4.22 & \starppp \\
    \midrule
     BSRI & {\scriptsize \textit{M}} &
       0.058 & \starppp & -1.98 &   & 5.71 & \starppp & 8.83 & \starppp & 
       \underline{21.233} & \starppp & 0.05 &   & 0.07 &   & 0.02 & \\
      & {\scriptsize \textit{F}} &
       0.037 & \starppp & -1.52 &   & 6.40 & \starppp & 8.56 & \starppp & 
       0.030 & \starppp & -3.93 & \starppp & -5.39 & \starppp & -1.75 & \\
    \midrule
     CABIN & {\scriptsize \textit{R}} &
       0.008 & \starp & 1.94 &   & 1.31 &   & -0.44 &   & 
       0.011 & \starp & -2.68 & \starp & -1.65 &   & 0.90 & \\
      & {\scriptsize \textit{I}} &
       0.007 &  & - &  & - &  & - &  & 
       0.016 & \starpp & -2.75 & \starp & 0.81 &   & 3.29 & \starpp \\
      & {\scriptsize \textit{A}} &
       0.009 & \starp & 2.81 & \starp & 1.93 &   & -0.85 &   & 
       0.010 & \starp & -1.95 &   & -0.20 &   & 1.74 & \\
      & {\scriptsize \textit{S}} &
       0.007 &  & - &  & - &  & - &  & 
       0.007 & \starp & -2.15 &   & 0.70 &   & 2.72 & \starp \\
      & {\scriptsize \textit{E}} &
       0.006 &  & - &  & - &  & - &  & 
       0.006 &  & - &  & - &  & - & \\
      & {\scriptsize \textit{C}} &
       0.017 & \starpp & 2.27 &   & 1.44 &   & -0.71 &   & 
       0.011 & \starp & -2.57 & \starp & 0.63 &   & 2.95 & \starp \\
    \midrule
     ICB & {\scriptsize \textit{O}} &
       0.020 & \starppp & -4.59 & \starppp & -2.37 &   & 1.68 &   & 
       0.012 & \starpp & -1.92 &   & -1.57 &   & 0.58 & \\
    \midrule
     ECR-R & {\scriptsize \textit{Anx.}} &
       0.003 &  & - &  & - &  & - &  & 
       0.109 & \starppp & -0.63 &   & -6.14 & \starppp & -6.85 & \starppp \\
      & {\scriptsize \textit{Avo.}} &
       0.022 & \starppp & -2.12 &   & 1.18 &   & 3.32 & \starpp & 
       0.104 & \starppp & -2.26 &   & -6.99 & \starppp & -5.59 & \starppp \\
\midrule
     MFQ-FF & {\scriptsize \textit{S. C}} &
       0.080 & \starppp & -4.76 & \starppp & -9.61 & \starppp & -4.83 & \starppp & 
       0.042 & \starppp & 6.15 & \starppp & 5.03 & \starppp & -1.43 & \\
      & {\scriptsize \textit{H}} &
       0.047 & \starppp & -4.79 & \starppp & -9.22 & \starppp & -4.52 & \starppp & 
       0.046 & \starppp & 6.32 & \starppp & 5.38 & \starppp & -1.45 & \\
      & {\scriptsize \textit{I}} &
       0.060 & \starppp & -4.79 & \starppp & -9.19 & \starppp & -4.39 & \starppp & 
       0.051 & \starppp & 6.17 & \starppp & 5.18 & \starppp & -1.43 & \\
      & {\scriptsize \textit{R}} &
       0.065 & \starppp & -4.46 & \starppp & -9.06 & \starppp & -4.61 & \starppp & 
       0.044 & \starppp & 5.97 & \starppp & 5.23 & \starppp & -1.11 & \\
      & {\scriptsize \textit{S-V}} &
       0.062 & \starppp & -4.72 & \starppp & -9.39 & \starppp & -4.67 & \starppp & 
       0.048 & \starppp & 6.10 & \starppp & 5.35 & \starppp & -1.08 & \\
      & {\scriptsize \textit{E}} &
       0.075 & \starppp & -4.67 & \starppp & -9.64 & \starppp & -4.97 & \starppp & 
       0.037 & \starppp & 5.87 & \starppp & 4.98 & \starppp & -1.33 & \\

    \midrule
     GSE & {\scriptsize \textit{O}} &
       0.001 &  & - &  & - &  & - &  & 
       0.001 &  & - &  & - &  & - & \\
    \midrule
     LOT-R & {\scriptsize \textit{O}} &
       0.084 & \starppp & -6.41 & \starppp & 3.76 & \starpp & 9.68 & \starppp & 
       0.020 & \starppp & -3.31 & \starpp & 1.55 &   & 4.74 & \starppp \\
    \midrule
     LMS & {\scriptsize \textit{R}} &
       0.006 & \starp & 0.06 &   & 2.96 & \starp & 3.19 & \starpp & 
       0.133 & \starppp & -6.63 & \starppp & -10.93 & \starppp & -4.59 & \starppp \\
      & {\scriptsize \textit{M}} &
       0.022 & \starppp & -4.73 & \starppp & -2.87 & \starp & 1.38 &   & 
       0.149 & \starppp & -5.97 & \starppp & -11.79 & \starppp & -6.26 & \starppp \\
      & {\scriptsize \textit{I}} &
       0.022 & \starppp & -5.09 & \starppp & -2.95 & \starp & 2.29 &   & 
       0.214 & \starppp & -7.76 & \starppp & -13.65 & \starppp & -7.41 & \starppp \\
    \midrule
     EIS & {\scriptsize \textit{O}} &
       0.027 & \starppp & -3.84 & \starppp & -0.63 &   & 3.21 & \starpp & 
       0.080 & \starppp & -1.55 &   & -5.55 & \starppp & -5.33 & \starppp \\
    \midrule
     WLEIS & {\scriptsize \textit{S}} &
       0.055 & \starppp & -3.17 & \starpp & 5.37 & \starppp & 9.04 & \starppp & 
       0.042 & \starppp & -4.89 & \starppp & -5.23 & \starppp & 0.17 & \\
      & {\scriptsize \textit{O}} &
       0.075 & \starppp & -4.21 & \starppp & 5.29 & \starppp & 9.67 & \starppp & 
       0.055 & \starppp & -5.49 & \starppp & -5.14 & \starppp & 0.75 & \\
      & {\scriptsize \textit{U}} &
       0.045 & \starppp & -4.08 & \starppp & 3.12 & \starpp & 7.33 & \starppp & 
       0.038 & \starppp & -5.14 & \starppp & -3.96 & \starppp & 1.65 & \\
      & {\scriptsize \textit{R}} &
       0.087 & \starppp & -3.26 & \starpp & 7.04 & \starppp & 11.19 & \starppp & 
       0.050 & \starppp & -5.44 & \starppp & -4.59 & \starppp & 1.79 & \\
    \midrule
     Empathy & {\scriptsize \textit{O}} &
       0.015 & \starppp & -2.59 & \starp & 1.58 &   & 4.53 & \starppp & 
       0.022 & \starppp & -1.74 &   & -3.49 & \starpp & -1.90 & \\
    \bottomrule
    \multicolumn{18}{r}{\starp $p<0.05$ \starpp $p<0.01$ \starppp $p<0.001$}
    \end{tabular}
    \caption{Result of statistical tests for GPT3.5-turbo and GPT4o. $Q$ columns indicate the Q-statistics from the Friedman test (except for GPT4o on BSRI Masculine factor, which shows F-statistics from ANOVA, marked with an underline). Also, $\Delta_{i,j}$ columns show the score difference between $i$-th and $j$-th snapshots and corresponding post-hoc test results.}
    \label{tab:app-full-1}
\end{table*}

\begin{table*}
    \footnotesize
    \centering
    \begin{tabular}{l@{}@{}r@{\,}|@{\,}r@{}l@{\,}r@{}l@{\,}r@{}l@{\,}r@{}l@{\,}|@{\,}r@{}l@{\,}r@{}l@{\,}r@{}l@{\,}r@{}l@{\,}|@{\,}r@{}l@{\,}r@{}l@{\,}r@{}l@{\,}r@{}l@{}}
    \toprule
        \multicolumn{2}{r|@{\,}}{\scriptsize \textit{Factors}} & 
        \multicolumn{8}{c|@{\,}}{LLaMA3.1 8B} & 
        \multicolumn{8}{c|@{\,}}{LLaMA3.1 70B} &
        \multicolumn{8}{c}{LLaMA3.1 405B} \\
        & & 
          $Q$ & & \multicolumn{2}{@{}c@{}}{$\Delta_{12,24}$} & \multicolumn{2}{@{}c@{}}{$\Delta_{24,36}$} & \multicolumn{2}{@{}c|@{\,}}{$\Delta_{12,36}$} &
          $Q$ & & \multicolumn{2}{@{}c@{}}{$\Delta_{12,24}$} & \multicolumn{2}{@{}c@{}}{$\Delta_{24,36}$} & \multicolumn{2}{@{}c|@{\,}}{$\Delta_{12,36}$} &
          $Q$ & & \multicolumn{2}{@{}c@{}}{$\Delta_{12,24}$} & \multicolumn{2}{@{}c@{}}{$\Delta_{24,36}$} & \multicolumn{2}{@{}c}{$\Delta_{12,36}$} \\
    \midrule
     BFI & {\scriptsize \textit{O}} &
0.021 & \starppp & 2.02 &   & 4.50 & \starppp & 3.08 & \starpp & 
0.004 &  & - &  & - &  & - &  & 
0.022 & \starppp & -0.16 &   & -2.88 & \starp & -3.22 & \starpp \\
      & {\scriptsize \textit{C}} &
0.036 & \starppp & 2.53 & \starp & 4.57 & \starppp & 2.31 &   & 
0.002 &  & - &  & - &  & - &  & 
0.030 & \starppp & -1.18 &   & -3.38 & \starpp & -2.73 & \starp \\
      & {\scriptsize \textit{E}} &
0.009 & \starp & -0.74 &   & 1.53 &   & 2.72 & \starp & 
0.011 & \starp & 0.75 &   & -2.01 &   & -3.68 & \starppp & 
0.010 & \starp & 0.00 &   & -1.83 &   & -2.10 &    \\
      & {\scriptsize \textit{A}} &
0.007 &  & - &  & - &  & - &  & 
0.004 &  & - &  & - &  & - &  & 
0.020 & \starppp & -0.52 &   & -3.16 & \starpp & -2.95 & \starp \\
      & {\scriptsize \textit{N}} &
0.010 & \starp & 2.51 & \starp & 3.50 & \starpp & 1.40 &   & 
0.006 &  & - &  & - &  & - &  & 
0.047 & \starppp & -1.63 &   & -4.98 & \starppp & -3.99 & \starppp \\
    \midrule
      EPQ-R & {\scriptsize \textit{E}} &
0.026 & \starppp & -2.37 &   & -4.19 & \starppp & -1.98 &   & 
0.017 & \starpp & -3.17 & \starpp & -6.13 & \starppp & -4.21 & \starppp & 
0.080 & \starppp & -3.75 & \starppp & -4.50 & \starppp & -1.84 & \\
      & {\scriptsize \textit{P}} &
0.033 & \starppp & -1.15 &   & -3.49 & \starpp & -2.55 & \starp & 
0.019 & \starppp & -1.12 &   & -3.79 & \starppp & -3.65 & \starppp & 
0.105 & \starppp & -3.93 & \starppp & -9.92 & \starppp & -7.23 & \starppp \\
      & {\scriptsize \textit{N}} &
0.023 & \starppp & -2.22 &   & -4.04 & \starppp & -2.22 &   & 
0.029 & \starppp & -1.63 &   & -4.94 & \starppp & -4.31 & \starppp & 
0.130 & \starppp & -3.87 & \starppp & -9.99 & \starppp & -7.27 & \starppp  \\
      & {\scriptsize \textit{L}} &
0.025 & \starppp & -1.21 &   & -4.27 & \starppp & -3.02 & \starpp & 
0.029 & \starppp & -0.59 &   & -4.61 & \starppp & -4.73 & \starppp & 
0.078 & \starppp & -2.94 & \starp & -8.63 & \starppp & -6.81 & \starppp \\
    \midrule
      DTDD & {\scriptsize \textit{M}} &
0.012 & \starpp & -4.08 & \starppp & -3.65 & \starppp & 0.28 &   & 
0.378 & \starppp & -12.97 & \starppp & -17.20 & \starppp & -6.50 & \starppp & 
0.121 & \starppp & -5.10 & \starppp & -8.82 & \starppp & -6.54 & \starppp \\
      & {\scriptsize \textit{P}} &
0.008 & \starp & -1.69 &   & -2.05 &   & -0.66 &   & 
0.426 & \starppp & -12.84 & \starppp & -18.08 & \starppp & -9.31 & \starppp & 
0.077 & \starppp & -3.40 & \starpp & -7.64 & \starppp & -6.03 & \starppp \\
      & {\scriptsize \textit{N}} &
0.004 &  & - &  & - &  & - &  & 
0.390 & \starppp & -12.28 & \starppp & -16.87 & \starppp & -8.50 & \starppp & 
0.051 & \starppp & -3.43 & \starpp & -6.33 & \starppp & -4.59 & \starppp \\
    \midrule
     BSRI & {\scriptsize \textit{M}} &
0.004 &  & - &  & - &  & - &  & 
0.051 & \starppp & -5.36 & \starppp & -7.96 & \starppp & -3.81 & \starppp & 
0.022 & \starppp & -3.93 & \starppp & -4.56 & \starppp & -1.12 &  \\
      & {\scriptsize \textit{F}} &
0.025 & \starppp & 4.19 & \starppp & 3.99 & \starppp & -0.23 &   & 
0.101 & \starppp & -3.54 & \starpp & -8.73 & \starppp & -6.09 & \starppp & 
0.019 & \starppp & -3.31 & \starpp & -3.77 & \starppp & -0.71 & \\
    \midrule
     CABIN & {\scriptsize \textit{R}} &
0.003 &  & - &  & - &  & - &  & 
0.099 & \starppp & 0.80 &   & -0.09 &   & -6.03 & \starppp & 
0.032 & \starppp & -2.15 &   & -4.30 & \starppp & -2.13 & \\
      & {\scriptsize \textit{I}} &
0.012 & \starpp & -0.83 &   & 0.23 &   & 1.01 &   & 
0.035 & \starppp & 2.20 &   & 0.09 &   & -2.95 & \starp & 
0.005 &  & - &  & - &  & - & \\
      & {\scriptsize \textit{A}} &
0.002 &  & - &  & - &  & - &  & 
0.052 & \starppp & -3.11 & \starpp & -5.75 & \starppp & -3.38 & \starpp & 
0.013 & \starpp & -2.22 &   & -3.54 & \starpp & -1.29 & \\
      & {\scriptsize \textit{S}} &
0.002 &  & - &  & - &  & - &  & 
0.065 & \starppp & -2.37 &   & -6.12 & \starppp & -4.56 & \starppp & 
0.022 & \starppp & -2.27 &   & -3.61 & \starpp & -1.32 & \\
      & {\scriptsize \textit{E}} &
0.003 &  & - &  & - &  & - &  & 
0.074 & \starppp & -3.32 & \starpp & -8.81 & \starppp & -6.11 & \starppp & 
0.034 & \starppp & -2.64 & \starp & -4.43 & \starppp & -1.40 &  \\
      & {\scriptsize \textit{C}} &
0.004 &  & - &  & - &  & - &  & 
0.117 & \starppp & -3.59 & \starpp & -9.47 & \starppp & -6.87 & \starppp & 
0.027 & \starppp & -3.20 & \starpp & -4.27 & \starppp & -0.86 & \\
    \midrule
     ICB & {\scriptsize \textit{O}} &
0.017 & \starpp & 2.73 & \starp & 3.03 & \starpp & 0.32 &   & 
0.018 & \starppp & 2.59 & \starp & 1.46 &   & -0.97 &   & 
0.016 & \starpp & -2.34 &   & -2.36 &   & -0.34 & \\
    \midrule
     ECR-R & {\scriptsize \textit{Anx.}} &
0.006 &  & - &  & - &  & - &  & 
0.092 & \starppp & -0.21 &   & -8.02 & \starppp & -8.40 & \starppp & 
0.124 & \starppp & 1.39 &   & -8.80 & \starppp & -11.05 & \starppp \\
      & {\scriptsize \textit{Avo.}} &
0.000 &  & - &  & - &  & - &  & 
0.086 & \starppp & 0.49 &   & -7.29 & \starppp & -7.87 & \starppp & 
0.110 & \starppp & 2.21 &   & -8.41 & \starppp & -10.21 & \starppp \\
\midrule
     MFQ-FF & {\scriptsize \textit{S. C}} &
0.004 &  & - &  & - &  & - &  & 
0.541 & \starppp & 15.53 & \starppp & 22.78 & \starppp & 12.07 & \starppp & 
0.207 & \starppp & 11.09 & \starppp & 12.99 & \starppp & 2.44 & \starp \\
      & {\scriptsize \textit{H}} &
0.002 &  & - &  & - &  & - &  & 
0.565 & \starppp & 15.50 & \starppp & 22.14 & \starppp & 11.51 & \starppp & 
0.302 & \starppp & 12.26 & \starppp & 15.40 & \starppp & 4.01 & \starppp \\
      & {\scriptsize \textit{I}} &
0.003 &  & - &  & - &  & - &  & 
0.550 & \starppp & 14.95 & \starppp & 21.51 & \starppp & 11.20 & \starppp & 
0.302 & \starppp & 12.63 & \starppp & 15.64 & \starppp & 3.50 & \starpp \\
      & {\scriptsize \textit{R}} &
0.003 &  & - &  & - &  & - &  & 
0.539 & \starppp & 14.75 & \starppp & 20.34 & \starppp & 10.52 & \starppp & 
0.263 & \starppp & 11.24 & \starppp & 13.55 & \starppp & 3.64 & \starppp \\
      & {\scriptsize \textit{S-V}} &
0.008 & \starp & -1.50 &   & -2.19 &   & -0.68 &   & 
0.564 & \starppp & 15.81 & \starppp & 22.14 & \starppp & 11.62 & \starppp & 
0.265 & \starppp & 12.33 & \starppp & 15.43 & \starppp & 3.69 & \starppp \\
      & {\scriptsize \textit{E}} &
0.007 &  & - &  & - &  & - &  & 
0.553 & \starppp & 15.55 & \starppp & 21.89 & \starppp & 11.40 & \starppp & 
0.273 & \starppp & 12.05 & \starppp & 14.83 & \starppp & 3.64 & \starppp  \\
    \midrule
     GSE & {\scriptsize \textit{O}} &
0.036 & \starppp & 3.52 & \starpp & 6.93 & \starppp & 3.90 & \starppp & 
0.126 & \starppp & 9.72 & \starppp & 4.19 & \starppp & -5.16 & \starppp & 
0.004 &  & - &  & - &  & - &  \\
    \midrule
     LOT-R & {\scriptsize \textit{O}} &
0.045 & \starppp & 3.93 & \starppp & 7.05 & \starppp & 3.83 & \starppp & 
0.027 & \starppp & 4.06 & \starppp & 1.18 &   & -0.65 &   & 
0.008 & \starp & 0.66 &   & 2.03 &   & 1.72 & \\
    \midrule
     LMS & {\scriptsize \textit{R}} &
0.004 &  & - &  & - &  & - &  & 
0.179 & \starppp & -5.79 & \starppp & -12.04 & \starppp & -9.44 & \starppp & 
0.268 & \starppp & -8.75 & \starppp & -15.46 & \starppp & -8.85 & \starppp  \\
      & {\scriptsize \textit{M}} &
0.023 & \starppp & 4.37 & \starppp & 3.89 & \starppp & -0.33 &   & 
0.169 & \starppp & -4.28 & \starppp & -11.10 & \starppp & -8.26 & \starppp & 
0.147 & \starppp & -7.36 & \starppp & -11.18 & \starppp & -5.62 & \starppp \\
      & {\scriptsize \textit{I}} &
0.020 & \starppp & 4.44 & \starppp & 4.36 & \starppp & 0.41 &   & 
0.215 & \starppp & -6.82 & \starppp & -12.96 & \starppp & -8.60 & \starppp & 
0.196 & \starppp & -5.57 & \starppp & -12.79 & \starppp & -7.98 & \starppp \\
    \midrule
     EIS & {\scriptsize \textit{O}} &
0.005 &  & - &  & - &  & - &  & 
0.277 & \starppp & -5.98 & \starppp & -12.73 & \starppp & -1.54 &   & 
0.105 & \starppp & -6.51 & \starppp & -9.34 & \starppp & -3.25 & \starpp \\
    \midrule
     WLEIS & {\scriptsize \textit{S}} &
0.003 &  & - &  & - &  & - &  & 
0.005 &  & - &  & - &  & - &  & 
0.034 & \starppp & -1.76 &   & 2.83 & \starp & 5.21 & \starppp \\
      & {\scriptsize \textit{O}} &
0.048 & \starppp & 5.18 & \starppp & 7.17 & \starppp & 2.45 & \starp & 
0.001 &  & - &  & - &  & - &  & 
0.013 & \starpp & -1.77 &   & 1.26 &   & 3.34 & \starpp \\
      & {\scriptsize \textit{U}} &
0.048 & \starppp & 5.64 & \starppp & 7.41 & \starppp & 2.36 &   & 
0.030 & \starppp & -2.06 &   & -4.09 & \starppp & -2.84 & \starp & 
0.022 & \starppp & 0.04 &   & 3.07 & \starpp & 3.23 & \starpp \\
      & {\scriptsize \textit{R}} &
0.044 & \starppp & 5.05 & \starppp & 7.30 & \starppp & 2.94 & \starp & 
0.011 & \starp & 1.23 &   & -1.60 &   & -3.03 & \starpp & 
0.006 &  & - &  & - &  & - & \\
    \midrule
     Empathy & {\scriptsize \textit{O}} &
0.001 &  & - &  & - &  & - &  & 
0.081 & \starppp & -0.81 &   & -7.01 & \starppp & -7.32 & \starppp & 
0.010 & \starp & 2.94 & \starp & 3.49 & \starpp & 1.14 & \\
    \bottomrule
    \multicolumn{26}{r}{\starp $p<0.05$ \starpp $p<0.01$ \starppp $p<0.001$}
    \end{tabular}
    \caption{Result of statistical tests for LLaMA3.1 model family. $Q$ columns indicate the Q-statistics from the Friedman test. Also, $\Delta_{i,j}$ columns show the score difference between $i$-th and $j$-th snapshots and corresponding post-hoc test results.}
    \label{tab:llama-comparison}
\end{table*}

\begin{table*}
    \footnotesize
    \centering
    \begin{tabular}{lr|r@{}lr@{}lr@{}lr@{}l|r@{}lr@{}lr@{}lr@{}l}
    \toprule
        \multicolumn{2}{r|}{\scriptsize \textit{Factors}} & 
        \multicolumn{8}{c|}{Mixtral 8x7B} & 
        \multicolumn{8}{c}{Mixtral 8x22B} \\
        & & 
          $Q$ & & \multicolumn{2}{c}{$\Delta_{12,24}$} & \multicolumn{2}{c}{$\Delta_{24,36}$} & \multicolumn{2}{c|}{$\Delta_{12,36}$} &
          $Q$ & & \multicolumn{2}{c}{$\Delta_{12,24}$} & \multicolumn{2}{c}{$\Delta_{24,36}$} & \multicolumn{2}{c}{$\Delta_{12,36}$} \\
    \midrule
     BFI & {\scriptsize \textit{O}} &
0.002 &  & - &  & - &  & - &  & 
0.012 & \starpp & -2.15 &   & -0.83 &   & -0.28 & \\
      & {\scriptsize \textit{C}} &
0.001 &  & - &  & - &  & - &  & 
0.010 & \starp & -1.16 &   & -0.98 &   & -0.67 & \\
      & {\scriptsize \textit{E}} &
0.003 &  & - &  & - &  & - &  & 
0.020 & \starppp & -3.63 & \starpp & -1.44 &   & -0.18 & \\
      & {\scriptsize \textit{A}} &
0.002 &  & - &  & - &  & - &  & 
0.004 &  & - &  & - &  & - & \\
      & {\scriptsize \textit{N}} &
0.007 &  & - &  & - &  & - &  & 
0.011 & \starp & -2.48 & \starp & -1.40 &   & -0.65 & \\
    \midrule
      EPQ-R & {\scriptsize \textit{E}} &
0.101 & \starppp & -3.22 & \starpp & -8.77 & \starppp & -6.95 & \starppp & 
0.025 & \starppp & -0.17 &   & -1.39 &   & -1.38 & \\
      & {\scriptsize \textit{P}} &
0.071 & \starppp & -2.21 &   & -8.19 & \starppp & -7.41 & \starppp & 
0.043 & \starppp & -1.51 &   & -1.41 &   & -1.32 & \\
      & {\scriptsize \textit{N}} &
0.110 & \starppp & -0.78 &   & -8.08 & \starppp & -8.44 & \starppp & 
0.034 & \starppp & 0.19 &   & -1.36 &   & -1.37 & \\
      & {\scriptsize \textit{L}} &
0.057 & \starppp & -1.60 &   & -7.33 & \starppp & -6.83 & \starppp & 
0.042 & \starppp & -0.80 &   & -1.41 &   & -1.37 & \\
    \midrule
      DTDD & {\scriptsize \textit{M}} &
0.013 & \starpp & -4.19 & \starppp & -3.78 & \starpp & -0.13 &   & 
0.018 & \starppp & -3.65 & \starppp & -3.83 & \starppp & -1.17 & \\
      & {\scriptsize \textit{P}} &
0.007 &  & - &  & - &  & - &  & 
0.010 & \starp & -2.61 & \starp & -3.34 & \starpp & -1.36 & \\
      & {\scriptsize \textit{N}} &
0.000 &  & - &  & - &  & - &  & 
0.009 & \starp & -1.46 &   & -2.80 & \starp & -1.63 & \\
    \midrule
     BSRI & {\scriptsize \textit{M}} &
0.002 &  & - &  & - &  & - &  & 
0.069 & \starppp & -2.84 & \starp & -3.70 & \starppp & -1.20 & \\
      & {\scriptsize \textit{F}} &
0.001 &  & - &  & - &  & - &  & 
0.065 & \starppp & -1.19 &   & -2.18 &   & -1.15 & \\
    \midrule
     CABIN & {\scriptsize \textit{R}} &
0.006 &  & - &  & - &  & - &  & 
0.015 & \starpp & 0.48 &   & -0.36 &   & -0.70 & \\
      & {\scriptsize \textit{I}} &
0.011 & \starp & -2.06 &   & -0.77 &   & 1.35 &   & 
0.003 &  & - &  & - &  & - & \\
      & {\scriptsize \textit{A}} &
0.011 & \starp & -2.04 &   & -0.70 &   & 1.40 &   & 
0.001 &  & - &  & - &  & - & \\
      & {\scriptsize \textit{S}} &
0.010 & \starp & -2.05 &   & -0.70 &   & 1.40 &   & 
0.001 &  & - &  & - &  & - & \\
      & {\scriptsize \textit{E}} &
0.006 &  & - &  & - &  & - &  & 
0.000 &  & - &  & - &  & - & \\
      & {\scriptsize \textit{C}} &
0.007 &  & - &  & - &  & - &  & 
0.002 &  & - &  & - &  & - & \\
    \midrule
     ICB & {\scriptsize \textit{O}} &
0.001 &  & - &  & - &  & - &  & 
0.002 &  & - &  & - &  & - & \\
    \midrule
     ECR-R & {\scriptsize \textit{Anx.}} &
0.033 & \starppp & 0.39 &   & -2.15 &   & -2.47 & \starp & 
0.085 & \starppp & -3.56 & \starpp & -5.75 & \starppp & -2.76 & \starp \\
      & {\scriptsize \textit{Avo.}} &
0.019 & \starppp & 0.17 &   & 0.54 &   & 0.29 &   & 
0.031 & \starppp & -1.24 &   & -2.06 &   & -0.95 & \\
    \midrule
MFQ-FF & {\scriptsize \textit{S. C}} &
0.004 &  & - &  & - &  & - &  & 
0.092 & \starppp & 3.08 & \starpp & 1.08 &   & -1.50 & \\
      & {\scriptsize \textit{H}} &
0.007 &  & - &  & - &  & - &  & 
0.103 & \starppp & 3.38 & \starpp & 1.65 &   & -1.43 & \\
      & {\scriptsize \textit{I}} &
0.006 &  & - &  & - &  & - &  & 
0.104 & \starppp & 3.41 & \starpp & 1.53 &   & -1.50 & \\
      & {\scriptsize \textit{R}} &
0.003 &  & - &  & - &  & - &  & 
0.109 & \starppp & 3.14 & \starpp & 1.48 &   & -1.32 & \\
      & {\scriptsize \textit{S-V}} &
0.005 &  & - &  & - &  & - &  & 
0.087 & \starppp & 3.58 & \starpp & 1.90 &   & -1.42 & \\
      & {\scriptsize \textit{E}} &
0.005 &  & - &  & - &  & - &  & 
0.094 & \starppp & 3.13 & \starpp & 1.59 &   & -1.29 & \\
    \midrule

     GSE & {\scriptsize \textit{O}} &
0.134 & \starppp & -9.93 & \starppp & -1.76 &   & 6.29 & \starppp & 
0.016 & \starpp & 0.89 &   & 0.05 &   & -0.50 & \\
    \midrule

     LOT-R & {\scriptsize \textit{O}} &
0.005 &  & - &  & - &  & - &  & 
0.013 & \starpp & 1.35 &   & 1.08 &   & 0.09 & \\
    \midrule
     LMS & {\scriptsize \textit{R}} &
0.081 & \starppp & -6.64 & \starppp & -7.86 & \starppp & -1.77 &   & 
0.037 & \starppp & -4.14 & \starppp & -4.57 & \starppp & -0.64 & \\
      & {\scriptsize \textit{M}} &
0.071 & \starppp & -4.83 & \starppp & -7.22 & \starppp & -2.43 & \starp & 
0.064 & \starppp & -4.73 & \starppp & -7.60 & \starppp & -2.82 & \starp \\
      & {\scriptsize \textit{I}} &
0.042 & \starppp & -3.89 & \starppp & -5.11 & \starppp & -1.38 &   & 
0.046 & \starppp & -4.92 & \starppp & -6.96 & \starppp & -2.64 & \starp \\
    \midrule
     EIS & {\scriptsize \textit{O}} &
0.061 & \starppp & -0.65 &   & -0.26 &   & 1.16 &   & 
0.020 & \starppp & -2.67 & \starp & -0.82 &   & 1.83 & \\
    \midrule
     WLEIS & {\scriptsize \textit{S}} &
0.000 &  & - &  & - &  & - &  & 
0.092 & \starppp & 5.44 & \starppp & 7.32 & \starppp & 2.45 & \starp \\
      & {\scriptsize \textit{O}} &
0.036 & \starppp & -0.73 &   & 4.10 & \starppp & 4.77 & \starppp & 
0.076 & \starppp & 5.02 & \starppp & 6.41 & \starppp & 1.09 & \\
      & {\scriptsize \textit{U}} &
0.027 & \starppp & -0.10 &   & 2.58 & \starp & 2.72 & \starp & 
0.071 & \starppp & 4.11 & \starppp & 4.55 & \starppp & 0.61 & \\
      & {\scriptsize \textit{R}} &
0.010 & \starp & -0.71 &   & 1.37 &   & 2.03 &   & 
0.087 & \starppp & 3.03 & \starpp & 2.53 & \starp & 0.04 & \\
    \midrule
     Empathy & {\scriptsize \textit{O}} &
0.021 & \starppp & -2.86 & \starp & -3.34 & \starpp & -1.15 &   & 
0.002 &  & - &  & - &  & - & \\
    \bottomrule
    \multicolumn{18}{r}{\starp $p<0.05$ \starpp $p<0.01$ \starppp $p<0.001$}
    \end{tabular}
    \caption{Result of statistical tests for Mixtral model family. $Q$ columns indicate the Q-statistics from the Friedman test. Also, $\Delta_{i,j}$ columns show the score difference between $i$-th and $j$-th snapshots and corresponding post-hoc test results.}
    \label{tab:mixtral-comparison}
\end{table*}

\begin{table*}
    \footnotesize
    \centering
    \begin{tabular}{lr|r@{}lr@{}lr@{}lr@{}l|r@{}lr@{}lr@{}lr@{}l}
    \toprule
        \multicolumn{2}{r|}{\scriptsize \textit{Factors}} & 
        \multicolumn{8}{c|}{Qwen2 7B} & 
        \multicolumn{8}{c}{Qwen2 72B} \\
        & & 
          $Q$ & & \multicolumn{2}{c}{$\Delta_{12,24}$} & \multicolumn{2}{c}{$\Delta_{24,36}$} & \multicolumn{2}{c|}{$\Delta_{12,36}$} &
          $Q$ & & \multicolumn{2}{c}{$\Delta_{12,24}$} & \multicolumn{2}{c}{$\Delta_{24,36}$} & \multicolumn{2}{c}{$\Delta_{12,36}$} \\
    \midrule
     BFI & {\scriptsize \textit{O}} &
0.016 & \starpp & -1.83 &   & -0.17 &   & 1.71 &   & 
0.010 & \starp & 1.26 &   & 2.61 & \starp & 1.73 & \\
      & {\scriptsize \textit{C}} &
0.007 & \starp & -1.84 &   & -0.06 &   & 1.78 &   & 
0.006 &  & - &  & - &  & - & \\
      & {\scriptsize \textit{E}} &
0.024 & \starppp & -1.27 &   & 0.49 &   & 1.54 &   & 
0.000 &  & - &  & - &  & - & \\
      & {\scriptsize \textit{A}} &
0.018 & \starppp & -1.69 &   & 0.11 &   & 1.73 &   & 
0.006 &  & - &  & - &  & - & \\
      & {\scriptsize \textit{N}} &
0.021 & \starppp & -1.82 &   & 0.00 &   & 1.80 &   & 
0.006 &  & - &  & - &  & - & \\
    \midrule
      EPQ-R & {\scriptsize \textit{E}} &
0.000 &  & - &  & - &  & - &  & 
0.003 &  & - &  & - &  & - & \\
      & {\scriptsize \textit{P}} &
0.002 &  & - &  & - &  & - &  & 
0.003 &  & - &  & - &  & - & \\
      & {\scriptsize \textit{N}} &
0.003 &  & - &  & - &  & - &  & 
0.004 &  & - &  & - &  & - & \\
      & {\scriptsize \textit{L}} &
0.003 &  & - &  & - &  & - &  & 
0.003 &  & - &  & - &  & - & \\
    \midrule
      DTDD & {\scriptsize \textit{M}} &
0.040 & \starppp & 3.50 & \starpp & 4.57 & \starppp & 1.24 &   & 
0.002 &  & - &  & - &  & - & \\
      & {\scriptsize \textit{P}} &
0.003 &  & - &  & - &  & - &  & 
0.003 &  & - &  & - &  & - & \\
      & {\scriptsize \textit{N}} &
0.000 &  & - &  & - &  & - &  & 
0.004 &  & - &  & - &  & - & \\
    \midrule
     BSRI & {\scriptsize \textit{M}} &
0.001 &  & - &  & - &  & - &  & 
0.002 &  & - &  & - &  & - & \\
      & {\scriptsize \textit{F}} &
0.005 &  & - &  & - &  & - &  & 
0.010 & \starp & -0.88 &   & 1.57 &   & 2.64 & \starp \\
    \midrule
     CABIN & {\scriptsize \textit{R}} &
0.028 & \starppp & -4.26 & \starppp & -4.70 & \starppp & -1.03 &   & 
0.027 & \starppp & -5.18 & \starppp & -2.87 & \starp & 2.45 & \starp \\
      & {\scriptsize \textit{I}} &
0.018 & \starppp & -3.54 & \starpp & -4.19 & \starppp & -1.03 &   & 
0.033 & \starppp & -5.30 & \starppp & -4.45 & \starppp & 1.16 & \\
      & {\scriptsize \textit{A}} &
0.021 & \starppp & -4.17 & \starppp & -4.34 & \starppp & -0.46 &   & 
0.046 & \starppp & -5.57 & \starppp & -4.65 & \starppp & 1.20 & \\
      & {\scriptsize \textit{S}} &
0.016 & \starpp & -4.06 & \starppp & -4.14 & \starppp & -0.35 &   & 
0.033 & \starppp & -4.32 & \starppp & -3.84 & \starppp & 0.54 & \\
      & {\scriptsize \textit{E}} &
0.023 & \starppp & -4.43 & \starppp & -4.39 & \starppp & -0.16 &   & 
0.022 & \starppp & -1.96 &   & -3.67 & \starppp & -1.13 & \\
      & {\scriptsize \textit{C}} &
0.020 & \starppp & -4.25 & \starppp & -4.26 & \starppp & -0.25 &   & 
0.017 & \starpp & -2.53 & \starp & -3.49 & \starpp & -0.63 & \\
    \midrule
     ICB & {\scriptsize \textit{O}} &
0.003 &  & - &  & - &  & - &  & 
0.036 & \starppp & 3.17 & \starpp & 3.40 & \starpp & 0.13 & \\
    \midrule
     ECR-R & {\scriptsize \textit{Anx.}} &
0.012 & \starpp & -0.92 &   & 2.49 & \starp & 3.70 & \starppp & 
0.003 &  & - &  & - &  & - & \\
      & {\scriptsize \textit{Avo.}} &
0.027 & \starppp & -4.55 & \starppp & -0.57 &   & 4.17 & \starppp & 
0.000 &  & - &  & - &  & - & \\
\midrule
     MFQ-FF & {\scriptsize \textit{S. C}} &
0.006 &  & - &  & - &  & - &  & 
0.108 & \starppp & 5.66 & \starppp & 8.55 & \starppp & 2.43 & \starp \\
      & {\scriptsize \textit{H}} &
0.002 &  & - &  & - &  & - &  & 
0.099 & \starppp & 5.79 & \starppp & 8.67 & \starppp & 2.46 & \starp \\
      & {\scriptsize \textit{I}} &
0.006 &  & - &  & - &  & - &  & 
0.105 & \starppp & 5.95 & \starppp & 8.50 & \starppp & 2.08 & \\
      & {\scriptsize \textit{R}} &
0.005 &  & - &  & - &  & - &  & 
0.100 & \starppp & 5.85 & \starppp & 8.73 & \starppp & 2.45 & \starp \\
      & {\scriptsize \textit{S-V}} &
0.004 &  & - &  & - &  & - &  & 
0.099 & \starppp & 5.75 & \starppp & 8.45 & \starppp & 2.30 & \\
      & {\scriptsize \textit{E}} &
0.009 & \starp & 3.46 & \starpp & 3.40 & \starpp & 0.16 &   & 
0.092 & \starppp & 5.80 & \starppp & 8.58 & \starppp & 2.38 & \\

    \midrule
     GSE & {\scriptsize \textit{O}} &
0.021 & \starppp & -3.48 & \starpp & 0.21 &   & 3.44 & \starpp & 
0.037 & \starppp & -2.35 &   & -2.57 & \starp & 1.03 & \\

    \midrule
     LOT-R & {\scriptsize \textit{O}} &
0.018 & \starppp & 3.56 & \starpp & 2.96 & \starpp & -0.45 &   & 
0.010 & \starp & 2.71 & \starp & 2.90 & \starp & 0.66 & \\
    \midrule
     LMS & {\scriptsize \textit{R}} &
0.065 & \starppp & -7.96 & \starppp & -4.88 & \starppp & 2.73 & \starp & 
0.006 &  & - &  & - &  & - & \\
      & {\scriptsize \textit{M}} &
0.022 & \starppp & -3.98 & \starppp & -2.02 &   & 1.92 &   & 
0.011 & \starp & 1.62 &   & 2.69 & \starp & 1.05 & \\
      & {\scriptsize \textit{I}} &
0.016 & \starpp & -2.82 & \starp & 0.41 &   & 3.35 & \starpp & 
0.003 &  & - &  & - &  & - & \\
    \midrule
     EIS & {\scriptsize \textit{O}} &
0.012 & \starpp & -4.10 & \starppp & -1.82 &   & 2.39 &   & 
0.048 & \starppp & -9.43 & \starppp & -8.32 & \starppp & 0.82 & \\
    \midrule
     WLEIS & {\scriptsize \textit{S}} &
0.084 & \starppp & -7.19 & \starppp & -5.68 & \starppp & 1.34 &   & 
0.011 & \starp & -3.00 & \starpp & 0.82 &   & 3.67 & \starpp \\
      & {\scriptsize \textit{O}} &
0.009 & \starp & -2.86 & \starp & -1.32 &   & 1.48 &   & 
0.024 & \starppp & -2.54 & \starp & 1.35 &   & 3.67 & \starpp \\
      & {\scriptsize \textit{U}} &
0.014 & \starpp & -1.80 &   & 1.38 &   & 3.26 & \starpp & 
0.061 & \starppp & -6.42 & \starppp & -2.66 & \starp & 3.67 & \starpp \\
      & {\scriptsize \textit{R}} &
0.036 & \starppp & -4.37 & \starppp & -1.20 &   & 3.48 & \starpp & 
0.014 & \starpp & -3.27 & \starpp & 0.07 &   & 3.42 & \starpp \\
    \midrule
     Empathy & {\scriptsize \textit{O}} &
0.003 &  & - &  & - &  & - &  & 
0.035 & \starppp & -2.69 & \starp & 2.87 & \starp & 5.72 & \starppp \\
    \bottomrule
    \multicolumn{18}{r}{\starp $p<0.05$ \starpp $p<0.01$ \starppp $p<0.001$}
    \end{tabular}
    \caption{Result of statistical tests for Qwen2 model family. $Q$ columns indicate the Q-statistics from the Friedman test. Also, $\Delta_{i,j}$ columns show the score difference between $i$-th and $j$-th snapshots and corresponding post-hoc test results.}
    \label{tab:qwen-comparison}
\end{table*}

\begin{table*}
    \footnotesize
    \centering
    \begin{tabular}{lr|r@{}lr@{}lr@{}lr@{}l|r@{}lr@{}lr@{}lr@{}l}
    \toprule
        \multicolumn{2}{r|}{\scriptsize \textit{Factors}} & 
        \multicolumn{8}{c|}{GPT4o-low} & 
        \multicolumn{8}{c}{GPT4o-high} \\
        & & 
          $Q$ & & \multicolumn{2}{c}{$\Delta_{12,24}$} & \multicolumn{2}{c}{$\Delta_{24,36}$} & \multicolumn{2}{c|}{$\Delta_{12,36}$} &
          $Q$ & & \multicolumn{2}{c}{$\Delta_{12,24}$} & \multicolumn{2}{c}{$\Delta_{24,36}$} & \multicolumn{2}{c}{$\Delta_{12,36}$} \\
    \midrule
     BFI & {\scriptsize \textit{O}} &
0.192 & \starppp & -6.06 & \starppp & -7.80 & \starppp & -2.97 & \starpp & 
0.099 & \starppp & -1.61 &   & -6.29 & \starppp & -5.06 & \starppp \\
      & {\scriptsize \textit{C}} &
0.106 & \starppp & -4.99 & \starppp & -5.36 & \starppp & -1.13 &   & 
0.063 & \starppp & -1.62 &   & -3.77 & \starppp & -2.76 & \starp \\
      & {\scriptsize \textit{E}} &
0.220 & \starppp & -6.79 & \starppp & -9.13 & \starppp & -3.38 & \starpp & 
0.051 & \starppp & -2.27 &   & -4.67 & \starppp & -2.29 &  \\
      & {\scriptsize \textit{A}} &
0.100 & \starppp & -5.47 & \starppp & -6.48 & \starppp & -1.76 &   & 
0.068 & \starppp & -3.75 & \starppp & -5.40 & \starppp & -1.92 &  \\
      & {\scriptsize \textit{N}} &
0.081 & \starppp & -3.62 & \starpp & -5.19 & \starppp & -1.78 &   & 
0.060 & \starppp & -2.82 & \starp & -3.98 & \starppp & -1.54 &  \\
    \midrule
      EPQ-R & {\scriptsize \textit{E}} &
0.283 & \starppp & -3.14 & \starpp & -10.28 & \starppp & -8.99 & \starppp & 
0.249 & \starppp & -2.42 & \starp & -9.25 & \starppp & -7.32 & \starppp  \\
      & {\scriptsize \textit{P}} &
0.283 & \starppp & -2.96 & \starp & -10.10 & \starppp & -9.02 & \starppp & 
0.299 & \starppp & -3.27 & \starpp & -10.18 & \starppp & -8.34 & \starppp \\
      & {\scriptsize \textit{N}} &
0.329 & \starppp & -3.79 & \starppp & -11.51 & \starppp & -9.63 & \starppp & 
0.273 & \starppp & -4.49 & \starppp & -10.61 & \starppp & -7.85 & \starppp \\
      & {\scriptsize \textit{L}} &
0.218 & \starppp & -2.34 &   & -9.60 & \starppp & -9.18 & \starppp & 
0.216 & \starppp & -2.46 & \starp & -9.34 & \starppp & -8.10 & \starppp \\
    \midrule
      DTDD & {\scriptsize \textit{M}} &
0.048 & \starppp & -4.56 & \starppp & -3.23 & \starpp & 0.52 &   & 
0.002 &  & - &  & - &  & - &  \\
      & {\scriptsize \textit{P}} &
0.055 & \starppp & -4.38 & \starppp & -4.29 & \starppp & -0.68 &   & 
0.001 &  & - &  & - &  & - & \\
      & {\scriptsize \textit{N}} &
0.029 & \starpp & -3.84 & \starppp & -3.08 & \starpp & 0.06 &   & 
0.008 &  & - &  & - &  & - &  \\
    \midrule
     BSRI & {\scriptsize \textit{M}} &
\underline{0.069} & \starppp & -6.60 & \starppp & -1.87 &   & 3.88 & \starppp & 
0.113 & \starppp & -5.34 & \starppp & -4.91 & \starppp & 0.21 & \\
      & {\scriptsize \textit{F}} &
0.082 & \starppp & -6.64 & \starppp & -3.05 & \starpp & 3.04 & \starpp & 
0.109 & \starppp & -5.76 & \starppp & -4.08 & \starppp & 1.04 & \\
    \midrule
     CABIN & {\scriptsize \textit{R}} &
0.110 & \starppp & -4.14 & \starppp & -6.40 & \starppp & -2.91 & \starp & 
0.078 & \starppp & -4.87 & \starppp & -8.16 & \starppp & -4.00 & \starppp \\
      & {\scriptsize \textit{I}} &
0.098 & \starppp & -3.51 & \starpp & -5.59 & \starppp & -3.22 & \starpp & 
0.086 & \starppp & -4.41 & \starppp & -7.75 & \starppp & -4.42 & \starppp \\
      & {\scriptsize \textit{A}} &
0.056 & \starppp & -3.76 & \starppp & -4.63 & \starppp & -1.44 &   & 
0.106 & \starppp & -4.30 & \starppp & -8.00 & \starppp & -4.14 & \starppp \\
      & {\scriptsize \textit{S}} &
0.092 & \starppp & -4.05 & \starppp & -6.37 & \starppp & -3.13 & \starpp & 
0.110 & \starppp & -4.70 & \starppp & -7.60 & \starppp & -3.72 & \starppp \\
      & {\scriptsize \textit{E}} &
0.081 & \starppp & -3.85 & \starppp & -5.63 & \starppp & -2.44 & \starp & 
0.117 & \starppp & -4.30 & \starppp & -8.44 & \starppp & -4.31 & \starppp \\
      & {\scriptsize \textit{C}} &
0.048 & \starppp & -3.39 & \starpp & -4.69 & \starppp & -1.75 &   & 
0.115 & \starppp & -4.95 & \starppp & -7.80 & \starppp & -3.11 & \starpp  \\
    \midrule
     ICB & {\scriptsize \textit{O}} &
0.025 & \starpp & -1.83 &   & -1.49 &   & 0.22 &   & 
0.073 & \starppp & -2.70 & \starp & -3.74 & \starppp & -1.34 &  \\
    \midrule
     ECR-R & {\scriptsize \textit{Anx.}} &
0.236 & \starppp & -3.82 & \starppp & -8.09 & \starppp & -5.33 & \starppp & 
0.064 & \starppp & 0.07 &   & -2.05 &   & -2.11 &  \\
      & {\scriptsize \textit{Avo.}} &
0.169 & \starppp & -3.22 & \starpp & -7.98 & \starppp & -4.61 & \starppp & 
0.007 &  & - &  & - &  & - &  \\
\midrule
     MFQ-FF & {\scriptsize \textit{S. C}} &
0.063 & \starppp & 4.81 & \starppp & 4.23 & \starppp & -1.09 &   & 
0.007 &  & - &  & - &  & - &  \\
      & {\scriptsize \textit{H}} &
0.067 & \starppp & 4.95 & \starppp & 4.24 & \starppp & -1.12 &   & 
0.010 &  & - &  & - &  & - & \\
      & {\scriptsize \textit{I}} &
0.071 & \starppp & 5.17 & \starppp & 4.41 & \starppp & -1.26 &   & 
0.007 &  & - &  & - &  & - & \\
      & {\scriptsize \textit{R}} &
0.060 & \starppp & 4.89 & \starppp & 4.43 & \starppp & -1.06 &   & 
0.005 &  & - &  & - &  & - & \\
      & {\scriptsize \textit{S-V}} &
0.074 & \starppp & 5.36 & \starppp & 4.53 & \starppp & -1.45 &   & 
0.006 &  & - &  & - &  & - & \\
      & {\scriptsize \textit{E}} &
0.058 & \starppp & 5.16 & \starppp & 4.52 & \starppp & -1.09 &   & 
0.007 &  & - &  & - &  & - & \\

    \midrule
     GSE & {\scriptsize \textit{O}} &
0.074 & \starppp & -1.55 &   & 4.57 & \starppp & 6.34 & \starppp & 
0.039 & \starppp & -3.94 & \starppp & -3.28 & \starpp & 0.47 & \\

    \midrule
     LOT-R & {\scriptsize \textit{O}} &
0.000 &  & - &  & - &  & - &  & 
0.051 & \starppp & -1.91 &   & -2.83 & \starp & -1.37 &  \\
    \midrule
     LMS & {\scriptsize \textit{R}} &
0.157 & \starppp & -5.85 & \starppp & -7.06 & \starppp & -2.70 & \starp & 
0.291 & \starppp & -8.11 & \starppp & -10.18 & \starppp & -4.89 & \starppp  \\
      & {\scriptsize \textit{M}} &
0.159 & \starppp & -7.23 & \starppp & -7.81 & \starppp & -2.43 & \starp & 
0.408 & \starppp & -8.66 & \starppp & -13.20 & \starppp & -7.26 & \starppp \\
      & {\scriptsize \textit{I}} &
0.196 & \starppp & -7.79 & \starppp & -8.42 & \starppp & -3.30 & \starpp & 
0.449 & \starppp & -9.87 & \starppp & -14.12 & \starppp & -8.18 & \starppp \\
    \midrule
     EIS & {\scriptsize \textit{O}} &
0.131 & \starppp & -6.93 & \starppp & -3.86 & \starppp & 2.62 & \starp & 
0.101 & \starppp & -4.84 & \starppp & -3.73 & \starppp & 0.88 & \\
    \midrule
     WLEIS & {\scriptsize \textit{S}} &
0.080 & \starppp & -5.28 & \starppp & -0.75 &   & 4.67 & \starppp & 
0.137 & \starppp & -5.33 & \starppp & -6.90 & \starppp & -2.22 &\\
      & {\scriptsize \textit{O}} &
0.021 & \starp & -2.95 & \starp & 0.14 &   & 2.87 & \starp & 
0.129 & \starppp & -5.96 & \starppp & -6.87 & \starppp & -1.03 & \\
      & {\scriptsize \textit{U}} &
0.073 & \starppp & -3.30 & \starpp & 1.35 &   & 5.17 & \starppp & 
0.095 & \starppp & -5.06 & \starppp & -6.40 & \starppp & -1.75 & \\
      & {\scriptsize \textit{R}} &
0.071 & \starppp & -3.03 & \starpp & 2.10 &   & 5.61 & \starppp & 
0.147 & \starppp & -6.14 & \starppp & -7.45 & \starppp & -1.47 & \\
    \midrule
     Empathy & {\scriptsize \textit{O}} &
0.042 & \starppp & -1.88 &   & -3.65 & \starpp & -1.99 &   & 
0.004 &  & - &  & - &  & - &  \\
    \bottomrule
    \multicolumn{18}{r}{\starp $p<0.05$ \starpp $p<0.01$ \starppp $p<0.001$}
    \end{tabular}
    \caption{Result of statistical tests for GPT4o-low and GPT4o-high. $Q$ columns indicate the Q-statistics from the Friedman test (except for GPT4o-low on BSRI Masculine factor, which shows F-statistics from ANOVA, marked with an underline). Also, $\Delta_{i,j}$ columns show the score difference between $i$-th and $j$-th snapshots and corresponding post-hoc test results.}
    \label{tab:gpt4o-lowhigh-comparison}
\end{table*}

\begin{table*}
    \footnotesize
    \centering
    \begin{tabular}{lr|r@{}lr@{}lr@{}lr@{}l|r@{}lr@{}lr@{}lr@{}l}
    \toprule
        \multicolumn{2}{r|}{\scriptsize \textit{Factors}} & 
        \multicolumn{8}{c|}{LLaMA3.1 405B-low} & 
        \multicolumn{8}{c}{LLaMA3.1 405B-high} \\
        & & 
          $Q$ & & \multicolumn{2}{c}{$\Delta_{12,24}$} & \multicolumn{2}{c}{$\Delta_{24,36}$} & \multicolumn{2}{c|}{$\Delta_{12,36}$} &
          $Q$ & & \multicolumn{2}{c}{$\Delta_{12,24}$} & \multicolumn{2}{c}{$\Delta_{24,36}$} & \multicolumn{2}{c}{$\Delta_{12,36}$} \\
    \midrule
     BFI & {\scriptsize \textit{O}} &
0.033 & \starpp & -1.88 &   & -2.60 & \starp & -1.25 &   & 
0.022 & \starp & -1.40 &   & -2.69 & \starp & -1.54 &  \\
      & {\scriptsize \textit{C}} &
0.016 & \starp & -1.61 &   & -2.30 &   & -1.32 &   & 
0.020 & \starp & -0.07 &   & -2.84 & \starp & -3.26 & \starpp \\
      & {\scriptsize \textit{E}} &
0.012 &  & - &  & - &  & - &  & 
0.019 & \starp & -0.48 &   & -3.05 & \starpp & -3.14 & \starpp \\
      & {\scriptsize \textit{A}} &
0.025 & \starpp & -1.98 &   & -3.06 & \starpp & -1.89 &   & 
0.034 & \starpp & -0.54 &   & -2.56 & \starp & -2.60 & \starp  \\
      & {\scriptsize \textit{N}} &
0.022 & \starp & -0.45 &   & -1.81 &   & -1.75 &   & 
0.021 & \starp & -0.86 &   & -2.18 &   & -1.72 & \\
    \midrule
      EPQ-R & {\scriptsize \textit{E}} &
0.125 & \starppp & 3.07 & \starpp & -3.57 & \starpp & -6.10 & \starppp & 
0.041 & \starppp & -0.84 &   & -3.91 & \starppp & -3.72 & \starppp \\
      & {\scriptsize \textit{P}} &
0.090 & \starppp & 2.37 &   & -4.42 & \starppp & -6.97 & \starppp & 
0.026 & \starpp & -0.90 &   & -4.77 & \starppp & -5.04 & \starppp \\
      & {\scriptsize \textit{N}} &
0.135 & \starppp & 2.48 & \starp & -5.01 & \starppp & -6.58 & \starppp & 
0.086 & \starppp & -1.15 &   & -5.58 & \starppp & -5.48 & \starppp\\
      & {\scriptsize \textit{L}} &
0.117 & \starppp & 2.29 &   & -4.98 & \starppp & -7.44 & \starppp & 
0.039 & \starppp & -1.30 &   & -4.29 & \starppp & -4.11 & \starppp  \\
    \midrule
      DTDD & {\scriptsize \textit{M}} &
0.006 &  & - &  & - &  & - &  & 
0.135 & \starppp & -4.91 & \starppp & -6.67 & \starppp & -3.73 & \starppp \\
      & {\scriptsize \textit{P}} &
0.007 &  & - &  & - &  & - &  & 
0.114 & \starppp & -3.82 & \starppp & -6.55 & \starppp & -4.07 & \starppp \\
      & {\scriptsize \textit{N}} &
0.017 & \starp & 3.43 & \starpp & 3.65 & \starpp & 1.21 &   & 
0.157 & \starppp & -1.92 &   & -7.14 & \starppp & -5.55 & \starppp  \\
    \midrule
     BSRI & {\scriptsize \textit{M}} &
0.024 & \starpp & -4.15 & \starppp & -1.72 &   & 2.17 &   & 
0.006 &  & - &  & - &  & - &  \\
      & {\scriptsize \textit{F}} &
0.040 & \starppp & -4.06 & \starppp & -2.63 & \starp & 1.48 &   & 
0.003 &  & - &  & - &  & - &  \\
    \midrule
     CABIN & {\scriptsize \textit{R}} &
0.008 &  & - &  & - &  & - &  & 
0.066 & \starppp & -3.47 & \starpp & -6.57 & \starppp & -3.65 & \starpp \\
      & {\scriptsize \textit{I}} &
0.006 &  & - &  & - &  & - &  & 
0.077 & \starppp & -3.06 & \starpp & -4.95 & \starppp & -2.33 & \\
      & {\scriptsize \textit{A}} &
0.002 &  & - &  & - &  & - &  & 
0.057 & \starppp & -3.28 & \starpp & -4.94 & \starppp & -1.92 &  \\
      & {\scriptsize \textit{S}} &
0.012 &  & - &  & - &  & - &  & 
0.059 & \starppp & -4.57 & \starppp & -6.36 & \starppp & -1.95 & \\
      & {\scriptsize \textit{E}} &
0.008 &  & - &  & - &  & - &  & 
0.063 & \starppp & -4.54 & \starppp & -5.91 & \starppp & -1.88 & \\
      & {\scriptsize \textit{C}} &
0.008 &  & - &  & - &  & - &  & 
0.082 & \starppp & -5.82 & \starppp & -5.55 & \starppp & -0.51 & \\
    \midrule
     ICB & {\scriptsize \textit{O}} &
0.003 &  & - &  & - &  & - &  & 
0.000 &  & - &  & - &  & - &  \\
    \midrule
     ECR-R & {\scriptsize \textit{Anx.}} &
0.088 & \starppp & 1.02 &   & -6.23 & \starppp & -7.88 & \starppp & 
0.091 & \starppp & 2.96 & \starp & -3.57 & \starpp & -7.08 & \starppp \\
      & {\scriptsize \textit{Avo.}} &
0.109 & \starppp & -0.12 &   & -7.35 & \starppp & -7.59 & \starppp & 
0.112 & \starppp & 2.05 &   & -5.12 & \starppp & -7.20 & \starppp  \\
\midrule
     MFQ-FF & {\scriptsize \textit{S. C}} &
0.448 & \starppp & 10.36 & \starppp & 11.67 & \starppp & 4.49 & \starppp & 
0.274 & \starppp & 3.46 & \starpp & 9.18 & \starppp & 5.82 & \starppp \\
      & {\scriptsize \textit{H}} &
0.502 & \starppp & 10.67 & \starppp & 13.32 & \starppp & 5.29 & \starppp & 
0.251 & \starppp & 3.45 & \starpp & 9.57 & \starppp & 6.32 & \starppp \\
      & {\scriptsize \textit{I}} &
0.571 & \starppp & 11.22 & \starppp & 13.11 & \starppp & 5.14 & \starppp & 
0.357 & \starppp & 4.22 & \starppp & 10.29 & \starppp & 5.91 & \starppp \\
      & {\scriptsize \textit{R}} &
0.400 & \starppp & 9.02 & \starppp & 11.35 & \starppp & 4.82 & \starppp & 
0.274 & \starppp & 4.45 & \starppp & 9.13 & \starppp & 5.77 & \starppp \\
      & {\scriptsize \textit{S-V}} &
0.490 & \starppp & 11.15 & \starppp & 12.88 & \starppp & 4.55 & \starppp & 
0.324 & \starppp & 4.27 & \starppp & 10.26 & \starppp & 6.02 & \starppp \\
      & {\scriptsize \textit{E}} &
0.440 & \starppp & 9.82 & \starppp & 11.75 & \starppp & 4.63 & \starppp & 
0.274 & \starppp & 3.60 & \starpp & 9.58 & \starppp & 5.10 & \starppp \\

    \midrule
     GSE & {\scriptsize \textit{O}} &
0.039 & \starppp & -1.81 &   & 3.54 & \starpp & 4.84 & \starppp & 
0.048 & \starppp & -1.88 &   & -4.01 & \starppp & -3.42 & \starpp \\

    \midrule
     LOT-R & {\scriptsize \textit{O}} &
0.025 & \starpp & 2.14 &   & 3.48 & \starpp & 1.82 &   & 
0.024 & \starpp & -0.21 &   & -2.32 &   & -2.47 & \starp \\
    \midrule
     LMS & {\scriptsize \textit{R}} &
0.029 & \starpp & -2.21 &   & -3.06 & \starpp & -1.45 &   & 
0.463 & \starppp & -5.34 & \starppp & -15.10 & \starppp & -12.07 & \starppp \\
      & {\scriptsize \textit{M}} &
0.005 &  & - &  & - &  & - &  & 
0.318 & \starppp & -4.01 & \starppp & -12.88 & \starppp & -9.92 & \starppp  \\
      & {\scriptsize \textit{I}} &
0.014 &  & - &  & - &  & - &  & 
0.270 & \starppp & -3.16 & \starpp & -11.08 & \starppp & -9.35 & \starppp \\
    \midrule
     EIS & {\scriptsize \textit{O}} &
0.132 & \starppp & -6.89 & \starppp & -5.78 & \starppp & 1.59 &   & 
0.011 &  & - &  & - &  & - &  \\
    \midrule
     WLEIS & {\scriptsize \textit{S}} &
0.056 & \starppp & 0.39 &   & 4.04 & \starppp & 3.54 & \starpp & 
0.005 &  & - &  & - &  & - & \\
      & {\scriptsize \textit{O}} &
0.025 & \starpp & -1.41 &   & 1.90 &   & 3.11 & \starpp & 
0.002 &  & - &  & - &  & - & \\
      & {\scriptsize \textit{U}} &
0.043 & \starppp & -2.41 & \starp & 1.73 &   & 3.56 & \starpp & 
0.001 &  & - &  & - &  & - & \\
      & {\scriptsize \textit{R}} &
0.018 & \starp & -1.05 &   & 2.09 &   & 2.78 & \starp & 
0.000 &  & - &  & - &  & - & \\
    \midrule
     Empathy & {\scriptsize \textit{O}} &
0.002 &  & - &  & - &  & - &  & 
0.002 &  & - &  & - &  & - & \\
    \bottomrule
    \multicolumn{18}{r}{\starp $p<0.05$ \starpp $p<0.01$ \starppp $p<0.001$}
    \end{tabular}
    \caption{Result of statistical tests for LLaMA3.1 405B-low and LLaMA3.1 405B-high. $Q$ columns indicate the Q-statistics from the Friedman test. Also, $\Delta_{i,j}$ columns show the score difference between $i$-th and $j$-th snapshots and corresponding post-hoc test results.}
    \label{tab:llama405b-lowhigh-comparison}
\end{table*}

\end{document}